\documentclass[10pt,conference]{IEEEtran}

\usepackage[utf8]{inputenc} 
\usepackage[T1]{fontenc}   
\usepackage{url}
\usepackage{booktabs}       
\usepackage{amsfonts}     
\usepackage{nicefrac}
\usepackage{xcolor}       
\usepackage{xspace}
\usepackage{fancyvrb}
\usepackage{fvextra}
\usepackage{float}
\usepackage{url}
\usepackage{booktabs}
\usepackage{xspace}
\usepackage{enumitem}
\usepackage[flushleft]{threeparttable}
\usepackage{makecell}
\usepackage{multirow}
\usepackage{xcolor, colortbl}
\usepackage{subcaption}
\usepackage{tikz}
\usepackage{wrapfig}
\usepackage{textcomp}
\usepackage{scalerel}  
\usepackage{tipa}
\usepackage{pifont}
\usepackage{multicol}
\usepackage[most]{tcolorbox}
\usepackage{listings}
\usepackage{caption}
\usepackage{lineno}
\usepackage{xurl}
\usepackage[hidelinks,breaklinks=true]{hyperref} %

\definecolor{purple5}{HTML}{7209B7}
\definecolor{purple4}{HTML}{580966}
\definecolor{purple3}{HTML}{770C8A}
\definecolor{purple2}{HTML}{9410AB}
\definecolor{purple1}{HTML}{C6259E}
\definecolor{pink}{HTML}{F83A90}
\definecolor{orange2}{HTML}{FA7543}
\definecolor{orange1}{HTML}{FC9D22}
\definecolor{yellow}{HTML}{FDC500}
\definecolor{rose}{HTML}{E91293}
\definecolor{lightpurple}{HTML}{8789C0}

\definecolor{darkblue}{rgb}{0, 0, 0.5}
\hypersetup{colorlinks=true, citecolor=purple4, linkcolor=purple4, urlcolor=purple4}

\newcommand{\Comment}[1]{}

\newcommand{\eg}[0]{e.g.,\xspace}
\newcommand{\ie}[0]{i.e.,\xspace}
\newcommand{\etal}[0]{et al.\xspace}
\newcommand{\cf}[0]{cf.\xspace}

\newcommand{\rqonea}[0]{RQ1a}
\newcommand{\rqoneb}[0]{RQ1b}
\newcommand{\rqonec}[0]{RQ1c}
\newcommand{\rqtwo}[0]{RQ2}
\newcommand{\rqthree}[0]{RQ3}

\newcommand{\mysec}[1]{\smallskip \noindent \textbf{\textit{#1.}}}

\newcommand{\CodeIn}[1]{{\small \texttt{#1}}} %

\newtcolorbox{keyInsightsBox}{enhanced, opacityback=0.75,colback=lightpurple!50!white,width=0.48\textwidth, colframe=white,arc=4mm,boxsep=2pt,    %
  top=3pt,       %
  bottom=3pt}

\newcommand*\circled[1]{\scalebox{0.8}{\tikz[baseline=(char.base)]{
\node[anchor=text, shape=circle,fill=lightpurple, inner sep=0pt, minimum size=1.2em] (char) {\footnotesize \textcolor{white}{#1}};}}}

\newcommand{\GH}{{\sc GitHub}\xspace}
\newcommand{\llm}{LLM\xspace}
\newcommand{\llmfull}{large language model\xspace}

\newcommand{\aicode}{agentic\xspace} %
\newcommand{\agentic}{agentic\xspace}
\newcommand{\humancode}{human\xspace} %

\newcommand{\aidev}{AIDev\xspace}
\newcommand{\agentsinthewild}{Agents in the Wild\xspace}
\newcommand{\semgrep}{Semgrep\xspace}
\newcommand{\osvscanner}{OSV-Scanner\xspace}
\newcommand{\sonarqube}{SonarQube\xspace}

\newcommand{\claudecodelogo}{\scalerel*{\includegraphics{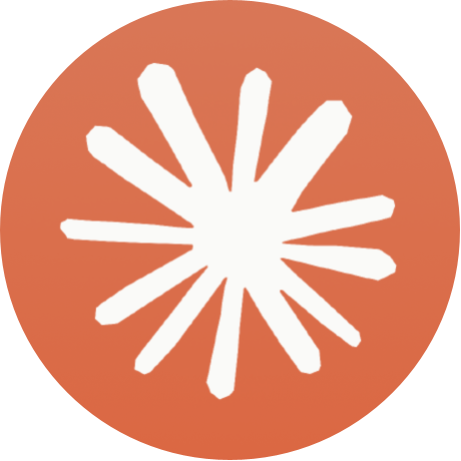}}{\textrm{C}}\xspace}
\newcommand{\codexlogo}{\scalerel*{\includegraphics{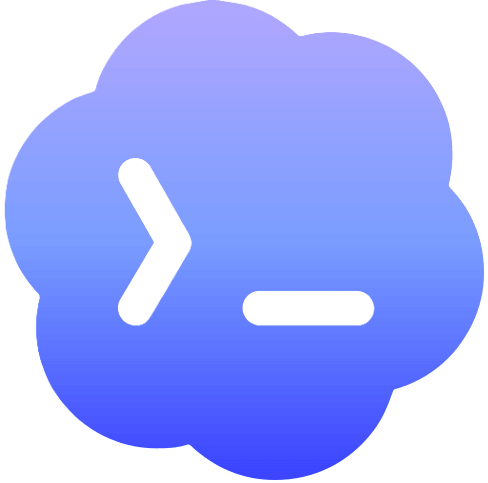}}{\textrm{C}}\xspace}
\newcommand{\copilotlogo}{\scalerel*{\includegraphics{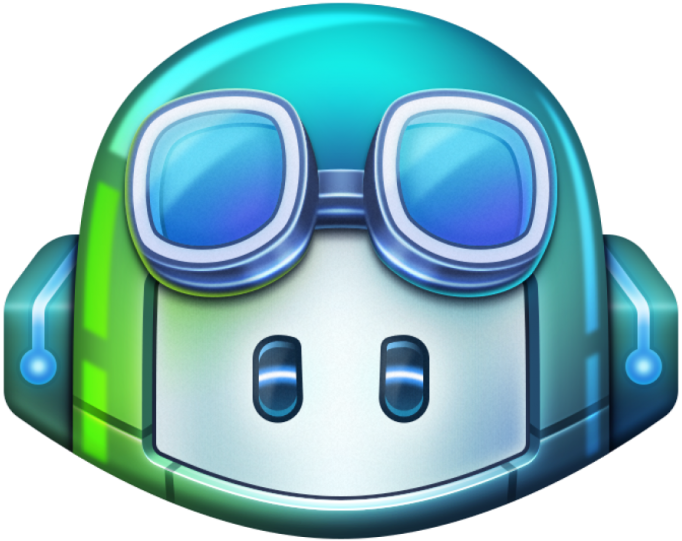}}{\textrm{C}}\xspace}
\newcommand{\cursorlogo}{\scalerel*{\includegraphics{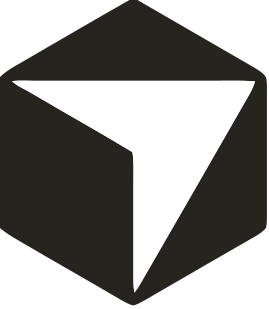}}{\textrm{C}}\xspace}
\newcommand{\devinlogo}{\scalerel*{\includegraphics{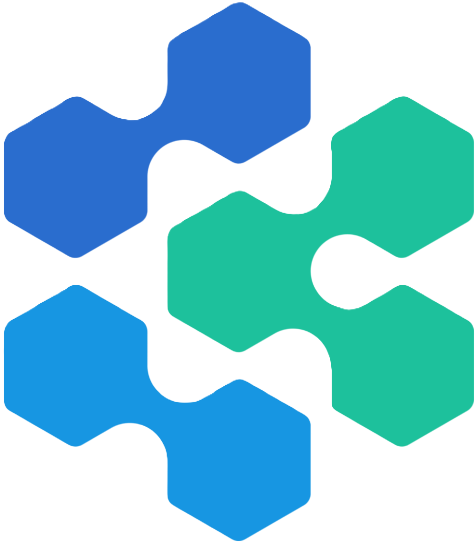}}{\textrm{C}}\xspace}
\newcommand{\aiderlogo}{\scalerel*{\includegraphics{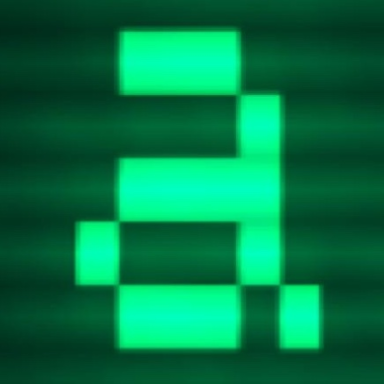}}{\textrm{C}}\xspace}
\newcommand{\openhandslogo}{\scalerel*{\includegraphics{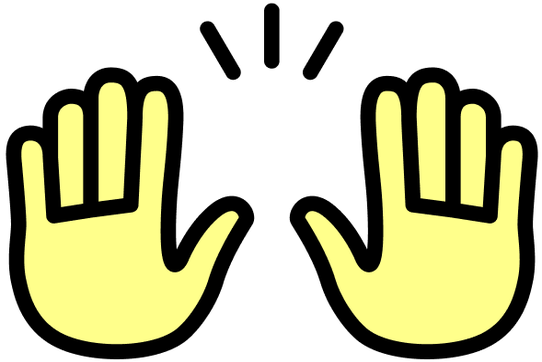}}{\textrm{C}}\xspace}
\newcommand{\clinelogo}{\scalerel*{\includegraphics{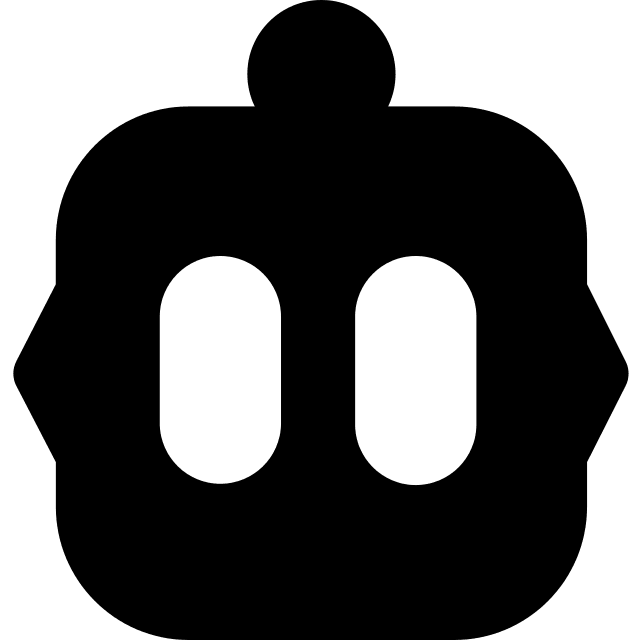}}{\textrm{C}}\xspace}
\newcommand{\roocodelogo}{\scalerel*{\includegraphics{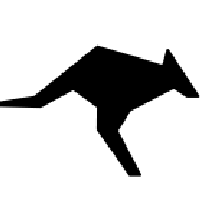}}{\textrm{C}}\xspace}
\newcommand{\windsurflogo}{\scalerel*{\includegraphics{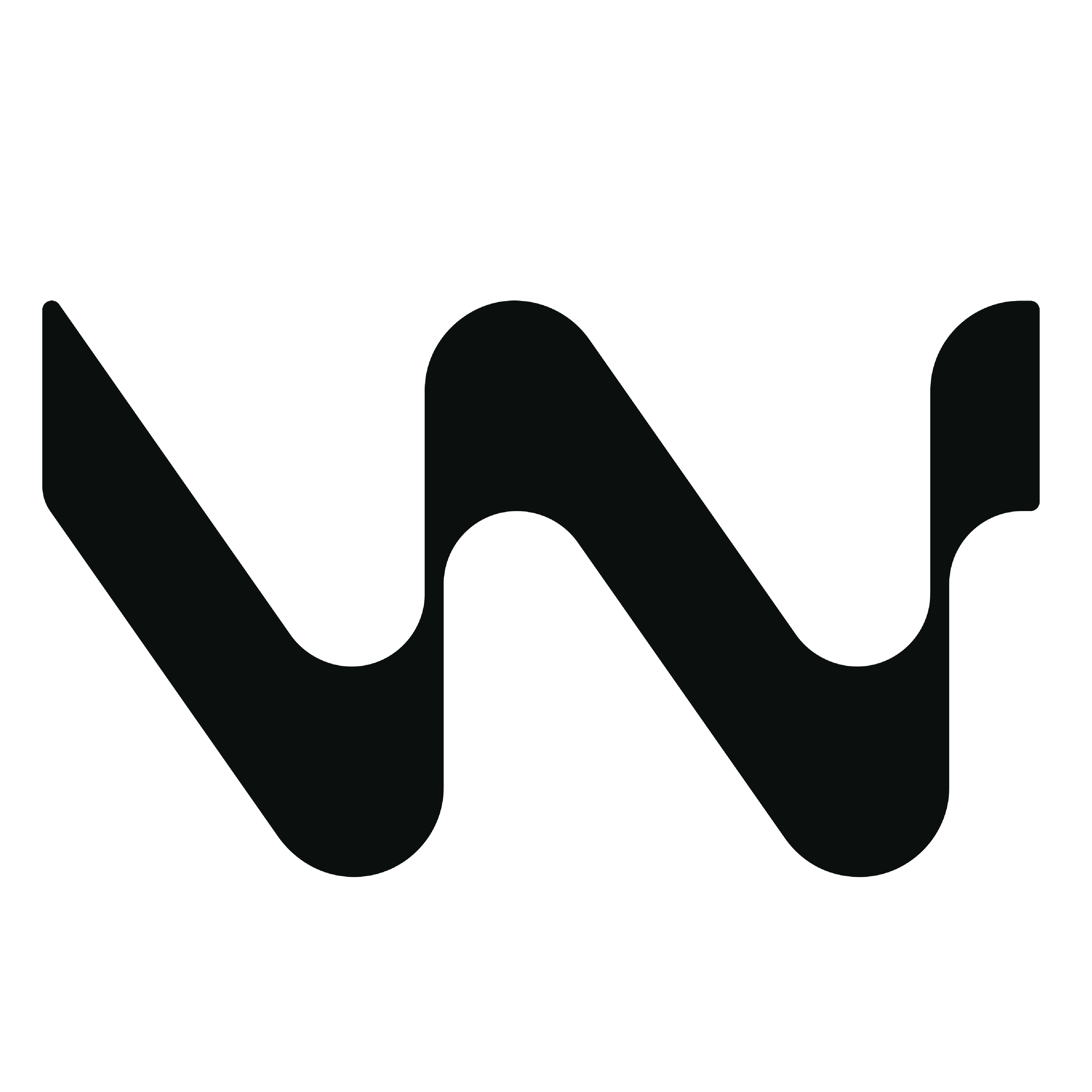}}{\textrm{C}}\xspace}
\newcommand{\sweeplogo}{\scalerel*{\includegraphics{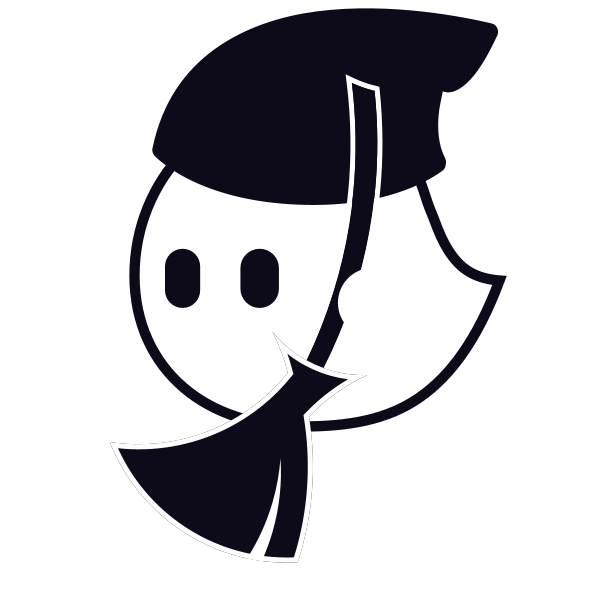}}{\textrm{C}}\xspace}
\newcommand{\replitlogo}{\scalerel*{\includegraphics{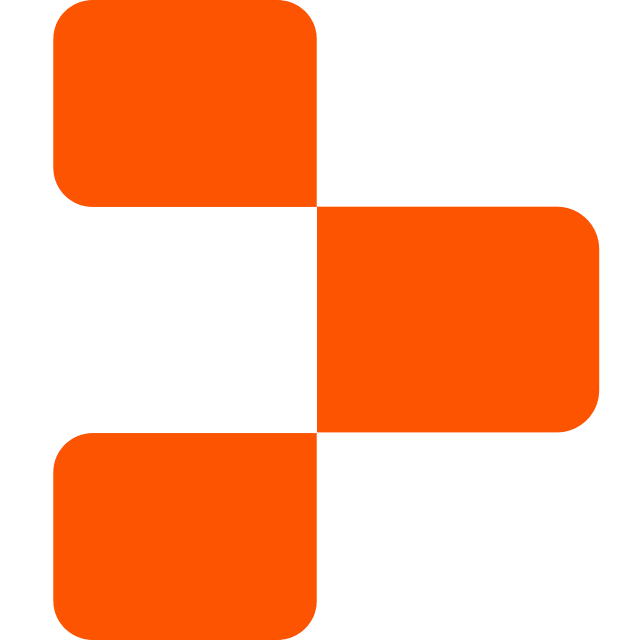}}{\textrm{C}}\xspace}
\newcommand{\juleslogo}{\scalerel*{\includegraphics{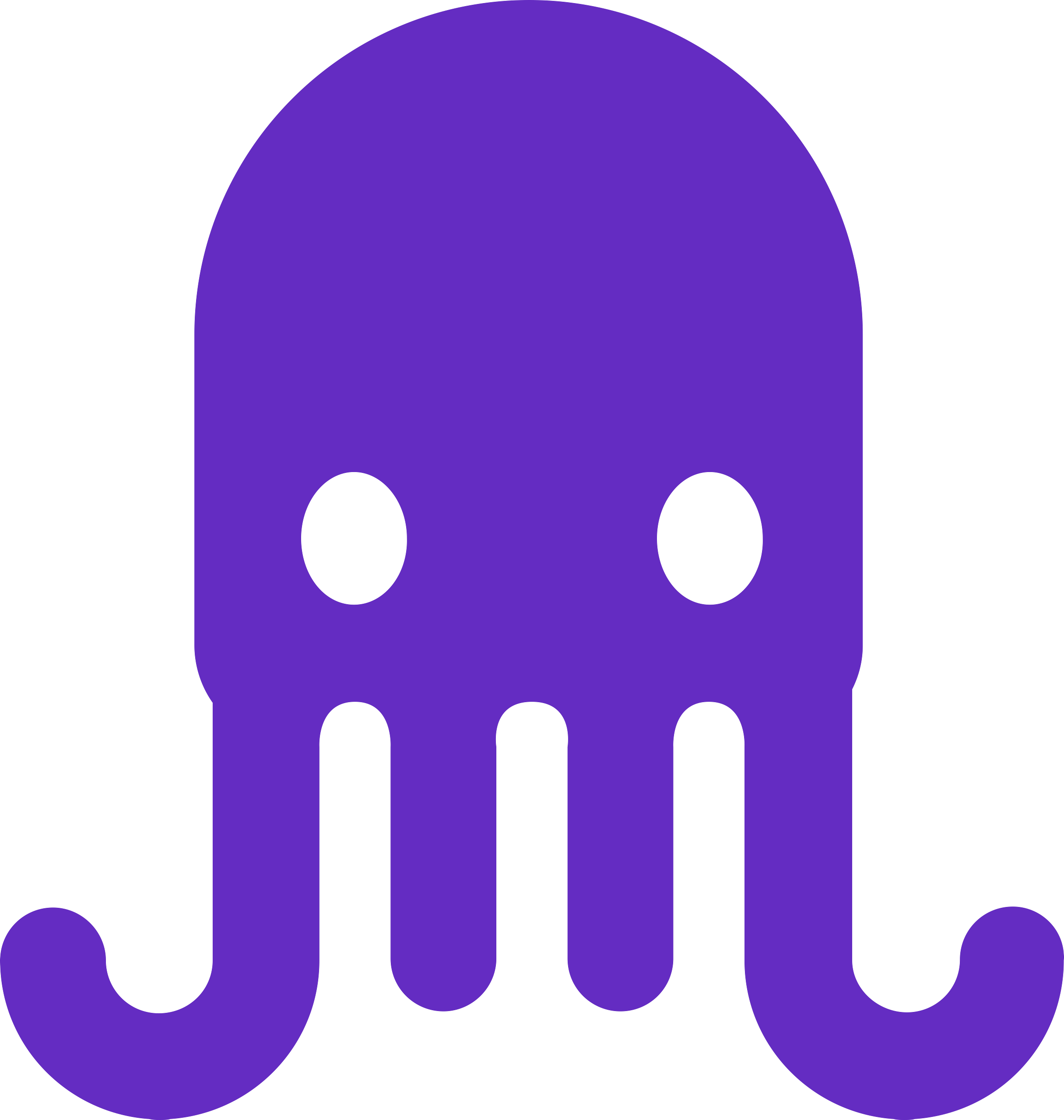}}{\textrm{C}}\xspace}
\newcommand{\geminilogo}{\scalerel*{\includegraphics{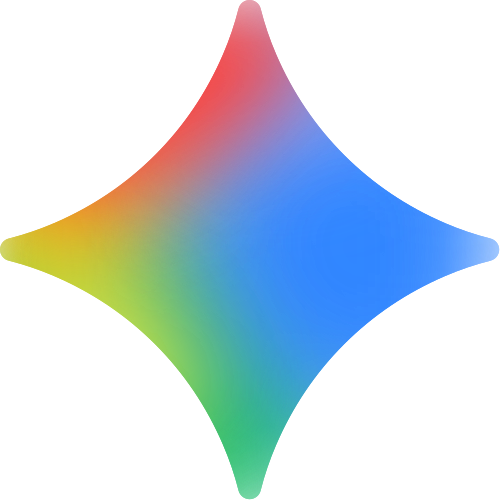}}{\textrm{C}}\xspace}
\newcommand{\amazonqlogo}{\scalerel*{\includegraphics{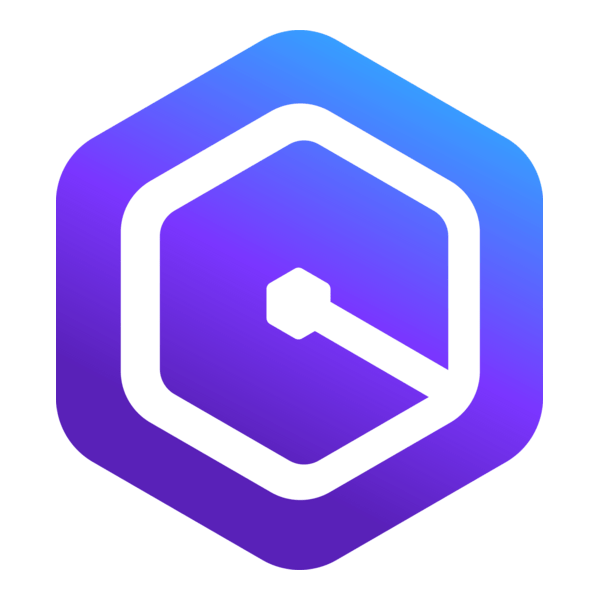}}{\textrm{C}}\xspace}

\newcommand{\uiuc}[1]{{#1\textsuperscript{\includegraphics[scale=0.004]{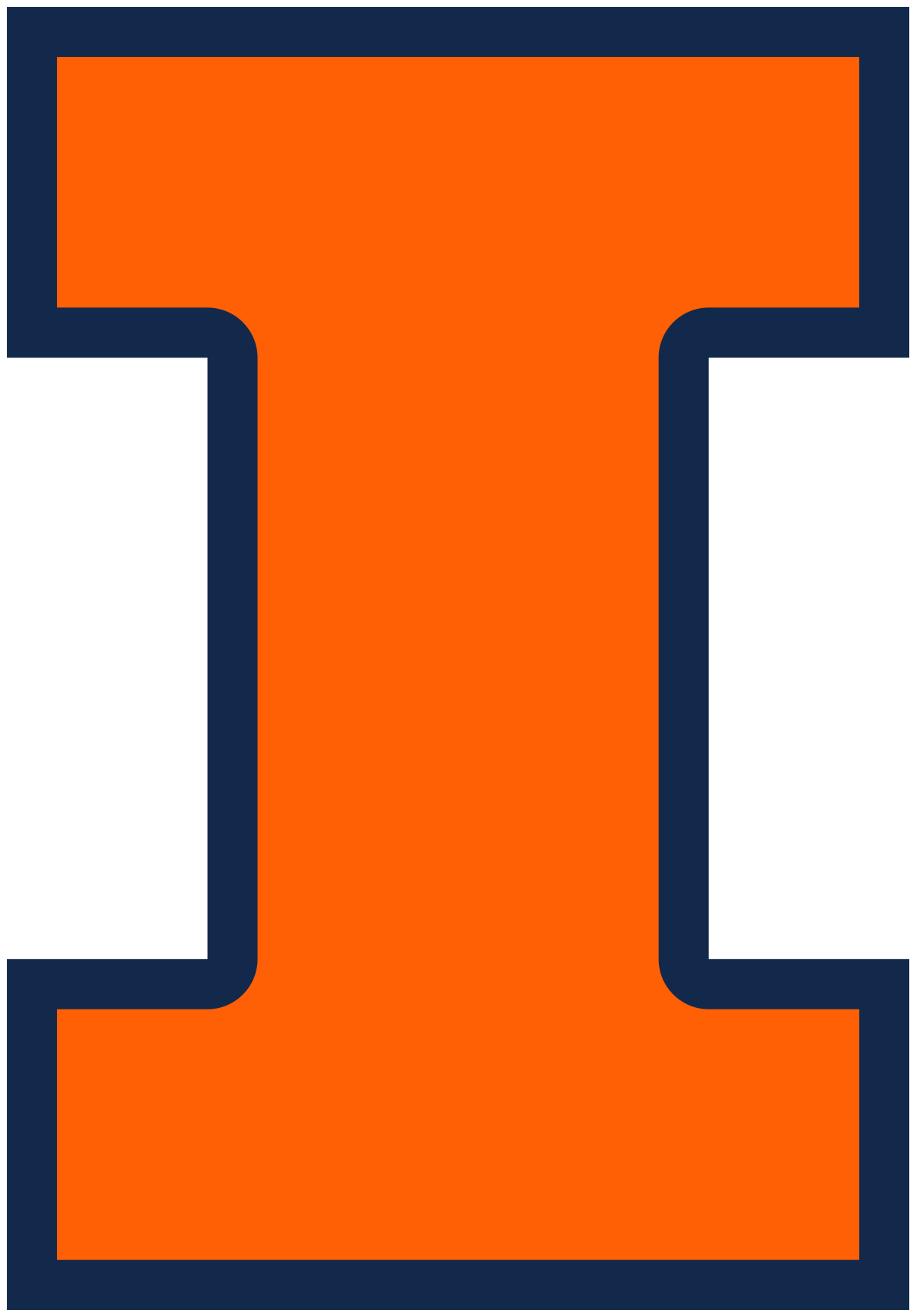}}}}
\newcommand{\gw}[1]{{#1\textsuperscript{\includegraphics[scale=0.064]{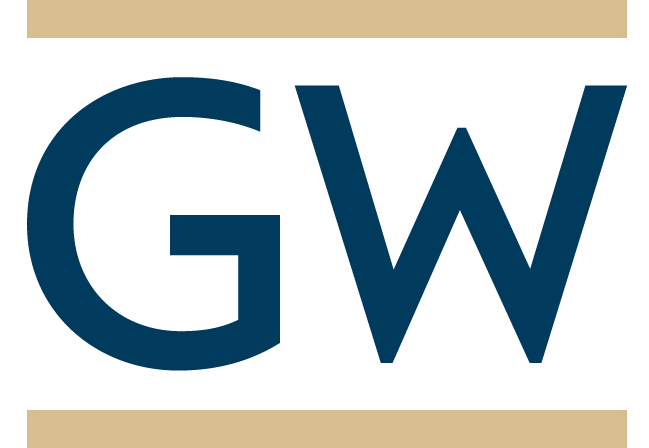}}}}

\author{
  
  \IEEEauthorblockN{Chunqiu Steven Xia}
  \IEEEauthorblockA{\uiuc{University of Illinois Urbana-Champaign}\\
  Urbana, IL, USA\\
  chunqiu2@illinois.edu}
  \and
  \IEEEauthorblockN{Courtney Miller}
  \IEEEauthorblockA{\gw{George Washington University}\\
  Washington, DC, USA\\
  courtney.miller@gwu.edu}

}

\begin{document}

\title{Do These Violent Delights Have Violent Ends? 
     \\Measuring the Post-Merge Fate of Agentic Code}

\maketitle

\frenchspacing %

\begin{abstract}
Agentic coding tools are increasingly used to make autonomous repository-level changes to real-world projects.
Prior work has largely evaluated these contributions at the pre-merge stage, through outcomes such as pull request acceptance and review effort.
Far less is known about what happens to \aicode code post-merge.
Yet merge success alone does not reveal whether a contribution will remain stable or require bug fixes and other corrective maintenance downstream. 
We conduct a longitudinal empirical analysis of \aicode and \humancode contributions across 182 repositories, tracking their post-merge fate over time, characterizing the intent of subsequent modifications, and analyzing the defects and vulnerabilities they introduce.
While the overall maintenance rates are similar, \aicode contributions require significantly higher rates of corrective maintenance and introduce more security weaknesses and dependency vulnerabilities.
We also find statistically significant evidence that \aicode maintenance burden is associated with repository characteristics.
In particular, each 10 percentage-point increase in a  project's no-review rate is associated with roughly a 6\% increase in \aicode maintenance burden on average. 
As coding agents become pervasive in software development, our findings highlight the need to evaluate and design agentic tools not only to produce mergeable changes, but to produce contributions that remain secure and maintainable.

\end{abstract}

\section{Introduction}
\label{sec:introduction}

The case for adopting Generative AI (GenAI) agentic coding tools is made almost entirely based on tool-centric metrics: how much the tools produce and what their output looks like at merge.
Microsoft reports GenAI produces close to 30\% of its codebase~\cite{novet25}, 
Meta aims for \aicode coding tools to handle half of its development by late 2026~\cite{mauran25}, 
and Google reports that 75\% of their new code is produced by GenAI~\cite{langley26}.
In a recent survey of more than 200 technology decision-makers, 67\% claim that \aicode coding tools write over half of their organization's weekly code~\cite{young26}.
These are the numbers used to justify adoption. 
They say little about that code's sustainability, reliability, security, or maintainability in the long term post-merge. 
What happens to \aicode code once it becomes part of the software system remains largely unstudied.

This tool-centric framing has revived evaluation metrics that are weak proxies for the quality of what actually ships: lines of code written, number of merged pull requests, percentage of code written by GenAI, and even ``tokenmaxxing" (amount of vendor tokens spent)~\cite{glover26}.
Such metrics make for impressive growth graphs, but they reveal little about how that change behaves as it evolves with the project.

\begin{figure}[t]
    \centering
    \includegraphics[width=\columnwidth]{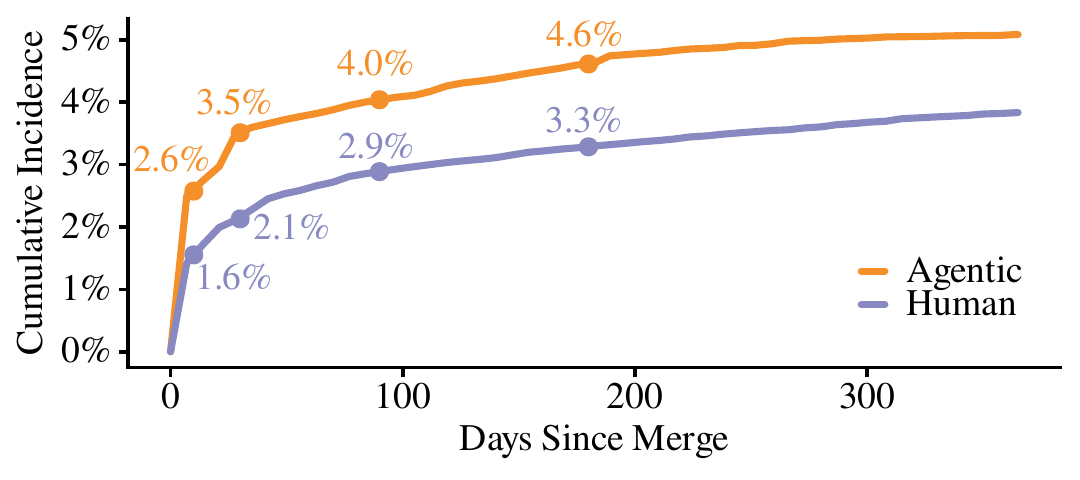}
    \caption{CIF for corrective terminations. Dots represent time points of 10/30/90/180 days since merge.
    }
    \label{fig:rq1b_cif_corrective}
\end{figure}

In fields like medicine and aviation, performance claims about tools that carry safety risks are expected to be substantiated before the tool is relied on. 
Medical devices like surgical robots must clear thorough regulatory review processes for safety and effectiveness before they can get FDA approval for use.
Agentic coding tools face no comparable bar before bold sweeping promises are made and they are deployed to write code for banks, hospitals, and other critical systems.
And whatever scrutiny the output receives largely occurs post-merge by the people who already paid for the tool and now potentially face another cost much greater. 
The actualized burden of this is already a common complaint among developers~\cite{gitlab26}, with many describing a dynamic where generation is cheap and then the arduous work of review and clean up falls on their shoulders~\cite{baltes26}. 
While practitioners have articulated this loudly and clearly, it has been argued largely from anecdote rather than measured in the code itself.

Almost everything we know about \aicode code quality comes from measurements taken at or before merge: performance on benchmarks~\cite{jimenez24}, pre-merge checks, and pull request acceptance rates and review outcomes~\cite{watanabe25, watanabe26}.
But merge only marks the beginning of a contribution's life in a project.
What happens to \aicode code after merge, and how it compares to \humancode code over time, remains largely unstudied.

\mysec{Our work} 
To bridge this gap, we present the first longitudinal empirical study of the post-merge lifecycle of \aicode and human code in real-world open source projects by exploring the following research questions (RQs):
\begin{enumerate}[leftmargin=26pt]
     \item[{\small \textbf{\rqonea}}] Does agentic code require more post-merge maintenance compared to human code?
     \item[{\small \textbf{\rqoneb}}] Does agentic code require different types of post-merge maintenance compared to human code?
     \item[{\small \textbf{\rqonec}}] Does agentic code require different types of corrective maintenance and introduce more defects compared to human code?
     \item[{\small \textbf{\rqtwo}}] Does corrective maintenance burden increase in projects as the proportion of agentic code increases over time? 
     \item[{\small \textbf{\rqthree}}] What project characteristics are associated with increased agentic maintenance burden?
 \end{enumerate}

We study 182 repositories with sustained agentic contributions, tracking what happens to \aicode and \humancode contributions over time.
We collect commits merged over a one-year window from May 1st 2025 to May 31st 2026. 
For each commit, we identify whether it was contributed by an \aicode coding tool or human, and we classify its maintenance purpose at two levels: a high-level \emph{maintenance class} (\eg corrective or perfective) and a finer-grained \emph{operational intent} (\eg bug fix, feature, or refactor). 
For each line in a commit, we perform a line-level lifecycle analysis that tracks whether the line survived or was terminated, when, and by which later commit, letting us connect every termination to the intent of the commit responsible for it. 
For each commit we also run \semgrep to identify source-code vulnerabilities and \osvscanner to identify dependency vulnerabilities, and we use \sonarqube to take weekly snapshots of each repository over the observation window. 
We then apply survival analysis, longitudinal panel models, and meta-regression to answer our RQs.

Our findings show that while \aicode code does not have a different maintenance burden in aggregate, that flat trend results from significant heterogeneity in the \aicode maintenance burden across projects. 
The composition of maintenance also differs: agentic code receives a 46\% higher corrective maintenance rate and a 45\% higher bug-fixing rate on average~(\cf Figure~\ref{fig:rq1b_cif_corrective}), and \aicode contributions introduce more security weaknesses and dependency vulnerabilities at merge. 
Agentic corrective maintenance burden also accumulates: within a project, a higher share of \aicode code in one month is associated with a higher corrective maintenance rate the following month, including corrective work performed by human developers.

Our model demonstrates the cross-project variation in maintenance burden is not random: a project's \aicode maintenance burden is associated with characteristics of the project itself.
Of the project characteristics we evaluate, a project's no-review merge rate has the strongest relationship with \aicode maintenance burden, 
with each 10 percentage-point increase in a  project's no-review rate being associated with a roughly 6\% increase in \aicode maintenance burden on average. 
Weaker test infrastructure and higher technical debt also both point in the same direction without reaching significance, together suggesting a picture where weaker engineering safeguards are associated with higher \aicode maintenance burden.

In summary, this paper makes the following contributions:
(1) A longitudinal dataset of \aicode and \humancode contributions across 182 repositories, linking each commit to its maintenance intent, line-level survival outcome, and static analysis findings.
(2) A comprehensive post-merge comparison of \aicode and \humancode contributions, covering overall maintenance activity, the composition and rate of corrective maintenance, the accumulation of corrective burden as \aicode code share grows over time, and some of the project characteristics associated with \aicode maintenance burden.

\section{Background and Related Work}
\label{sec:relatedwork}

\subsection{\llm{s} and Coding Agents}

With the recent rise of \llmfull{s} (\llm{s}) and their high benchmark performance on software engineering tasks~\cite{chen21}, developers have started adopting \llm{s} as part of their development workflows.
Initially, \emph{\llm-based coding assistants} such as GitHub Copilot~\cite{copilot} and Cursor~\cite{cursor} were developed to be integrated into developers' IDEs.
These coding assistants provide inline code completion suggestions based on the surrounding code context which users could accept, modify, or reject.
More recently, \emph{autonomous coding agents} have emerged to move beyond inline completion to perform repository-level development tasks, \eg feature implementation, refactoring, code execution, and pull request submission.
Tools like Anthropic's \claudecodelogo Claude Code~\cite{claudecode} and OpenAI's \codexlogo{} Codex~\cite{codex} have access to the full codebase with a chat-based interface that allows the user to interact and collaborate to perform development tasks.
In this sense, the usage of coding agents shifts the developer's role from manual coding toward high-level task orchestration with the promise of drastically improving developer efficiency~\cite{miller26b}.
\GH reports more than 1 million open-source repositories used GenAI coding tools between 2024 and 2025~\cite{github25b}, and OpenAI reports more than 5 million weekly Codex users as of June 2026~\cite{openai26}, with a similar number of users being estimated for Claude Code~\cite{bort26}.
These numbers suggest the adoption of coding agents shows no sign of slowing down.
However, despite this rapid adoption and the promise of improved developer productivity, our understanding of how coding agents affect software development remains limited.
Questions about how developers interact with \aicode contributions, how those contributions differ from human code, and what longitudinal maintenance, correctness, and security risks they introduce must be continually examined.

\subsection{Coding Agents Impact on Software Development}

Driven by the recent adoption of GenAI-based tools, researchers have begun evaluating their impact on software development.
Initial research mainly focuses on \llm-based coding assistants across areas of code quality~\cite{nguyen22, yetistiren22, borek25}, code security~\cite{asare23, pearce22, perry23}, maintainability~\cite{borg26, paul25}, and usability and productivity~\cite{chretien24, ng24, paradis25, shihab25, ziegler22, vaithilingam22, imai22, peng23}.
These studies report mixed productivity effects, ranging from moderate gains~\cite{peng23, song24, cui26, pandey24} to measurable decreases~\cite{becker25, xu25}.
At the same time, growing evidence raises concerns about the trustworthiness of generated code, including security vulnerabilities~\cite{ambati24, asare23}, regressions~\cite{li24b}, code smells~\cite{siddiq22}, and outdated API usage~\cite{kharma26}.

More recently, following the adoption trend, researchers have begun studying the impact of autonomous coding agents on software development~\cite{ogenrwot26, liu26}.
Watanabe~\etal~\cite{watanabe25} found that more than 80\% of Claude Code pull requests were eventually merged, with less than 50\% of agentic pull requests requiring additional human revisions.
He~\etal~\cite{he26} studied the impact of Cursor adoption across 807 \GH repositories and found that although project development velocity increased temporarily, code complexity and static analysis warnings also increased persistently, ultimately slowing project velocity in the long term.
Researchers have also begun collecting datasets of real-world \aicode coding tool usage, including \aidev~\cite{li25}, which captures more than 450,000 \aicode pull requests, and \agentsinthewild~\cite{murtz25}, which continuously tracks \aicode pull requests on \GH.
Using and building on these datasets, recent studies have examined agentic pull-request outcomes and failure modes~\cite{peralta26, nakashima26}, review dynamics~\cite{haider26, minh26, watanabe26}, code change characteristics~\cite{haque26, horikawa25}, and repository-level outcomes~\cite{agarwal26, cynthia26}.
Taken together, these studies show that while autonomous coding agents can produce mergeable pull requests at a high rate, they may also strain the review process and increase project complexity.

However, most existing work focuses on the pre-merge stage. 
A few studies include brief post-merge analyses, but only through narrow windows.
Rahman and Shihab~\cite{rahman26} focus purely on \aicode code survival and use only coarse modification intent while Liu et al.~\cite{liu26} focuses on the persistence of static-analysis issues introduced by \aicode commits.
Both provide post-merge signals, but neither reconstructs the broader maintenance trajectory of \aicode contributions.
Our study addresses this gap by following \aicode and \humancode code after merge, identifying the later commits that modify them, characterizing the intent of those changes, measuring defects and vulnerabilities introduced, and studying how \aicode code share and project characteristics affect maintenance burden.

\section{Methods}
\label{sec:methods}

\begin{table}[t]
  \centering
  \begin{tabular}{lrrr}
\toprule
Metric & Mean & Min & Max \\
\midrule
Stars & 4,867.2 & 10 & 45,401 \\
Forks & 945.6 & 0 & 15,123 \\
Human contributors & 70.1 & 1 & 443 \\
Unique AI agents used & 2.4 & 1 & 5 \\
Commits & 2,273.5 & 153 & 19,934 \\
AI-coauthored commits & 169.8 & 1 & 2,397 \\
Human commits & 1,860.7 & 42 & 18,711 \\
\bottomrule
\end{tabular}

  \caption{Descriptive statistics of our dataset.}
  \label{tab:repo_overall_stats}
\end{table}

\begin{figure*}
    \centering
    \includegraphics[width=\linewidth]{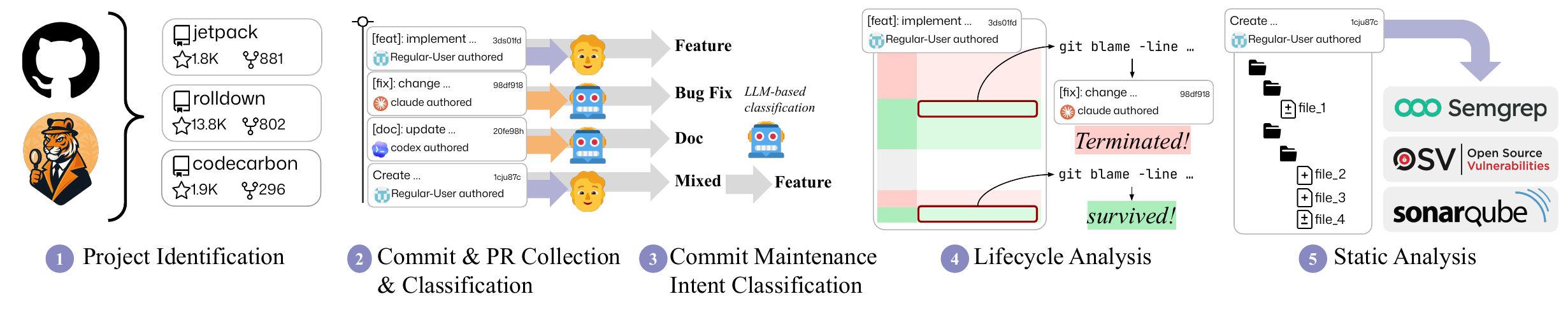}
    \caption{Overview of the data collection process.}
    \label{fig:overview}
\end{figure*}

Table~\ref{tab:repo_overall_stats} shows the overall summary characteristics of the repositories in our dataset.
We conduct our longitudinal analysis over an observation window from May 1st 2025 to May 31st 2026. 
We start at May 2025 because that is when we observe a meaningful increase in \aicode contributions.
Figure~\ref{fig:overview} shows the overview of our data collection approach. 

\mysec{\textnormal{\circled{1}} Project Identification}
We start by identifying projects with \aicode contributions. 
We use the \agentsinthewild dataset~\cite{murtz25} which contains pull requests (PRs) made by AI-coding agents on public \GH repositories. 
This yields an initial set of 2,860 projects with at least 10 stars and at least one human and one agentic PR.
We remove 29 projects from the top and bottom 1\% of the star distribution to reduce the influence of extreme popularity outliers.
Next, to filter out projects without sustained agentic activity, we exclude 2,631 projects without at least 10 agentic and 10 human PRs overall, at least three months of agentic PRs, and at least 10 commits per month over the past six months.
We further filter out one non-source-code project.
Lastly, we remove 17 extremely large projects, whose size hinders our lifecycle and static analysis to end with 182 total projects in our dataset.

\begin{table}[t]
  \centering
  \begin{tabular}{l}
\toprule
Agents considered \\
\midrule
\makecell[l]{\claudecodelogo Claude Code, \codexlogo Codex, \copilotlogo GitHub Copilot, \cursorlogo Cursor, \devinlogo Devin,\\
\aiderlogo Aider, \openhandslogo OpenHands, \clinelogo Cline, \roocodelogo Roo Code, \windsurflogo Windsurf, \\
\sweeplogo Sweep, \replitlogo Replit Agent, \juleslogo Jules, \geminilogo Gemini, \amazonqlogo Amazon Q} \\
\bottomrule
\end{tabular}

  \caption{GenAI coding agents targeted in study.}
  \label{tab:commit_author_classification}
\end{table}

\mysec{\textnormal{\circled{2}} Commit \& PR Collection \& Classification}
For each project in our dataset, we collect all commits merged during our observation window.
To identify \aicode commits, we use a simple classification procedure that matches commit author name, email, and commit signature against known agent names (\eg{} \CodeIn{Co-authored-by: Claude <noreply@anthropic.com>}).
Table~\ref{tab:commit_author_classification} lists the AI coding agents that we specifically target.
In addition, we collect basic PR review metrics including the number of comments and review threads.

\begin{table}[t]
  \centering
  \begin{tabular}{ll}
\toprule
Maintenance Class & Operational Intent \\
\midrule
Corrective & Security fix, Revert, Bug fix \\
Adaptive & Dependency update \\
Perfective & Performance, Feature, Documentation, Resource \\
Preventive & Refactor, Test, Style/Formatting \\
Management & Merge/Release/Versioning, Build/Config/CI \\
\bottomrule
\end{tabular}

  \caption{Two levels of maintenance intent categories.}
  \label{tab:commit_intent}
\end{table}

\mysec{\textnormal{\circled{3}} Commit Maintenance Intent Classification}
We classify the maintenance intent of each commit at two levels (\cf~Table~\ref{tab:commit_intent}): (1) high-level \textit{maintenance class} and (2) low-level \textit{operational intent}.
For maintenance class, we extend the classic Swanson's taxonomy~\cite{swanson76, varga18} with an additional \textit{management} category for commits related to merging, release, and versioning which are not captured by the original taxonomy. 
For operational intent, we borrow concepts from prior large-scale study~\cite{hindle08, wang21} as classification signals.

We first classify each commit into the low-level operational intent and then map that intent to the high-level maintenance class according to Table~\ref{tab:commit_intent}.
We begin by tagging the modified files by file types (\textit{source code}, \textit{test}, \textit{doc}, \textit{config}, \textit{dependency}, and \textit{resource}) based on file extensions and paths.
We then use these file-type tags to assign scores to specific intents.
For example, a commit that modifies only test files will have scores added to the test intent.
Next, we analyze the commit message subject by first checking for conventional commit prefixes~\cite{zeng25} and mapping them to operational intents.
We then apply keyword{ regexes} commonly associated with each operational intent on the commit subject text.
Finally, we obtain a score distribution over all operational intents and determine a primary intent by selecting the intent with the highest score.
We also categorize commits as \textit{mixed} if there are multiple intents with the same score or as \textit{unknown} if no intent receives a sufficiently high score.

To handle unknown or mixed commits, we perform an \llm-based reclassification.
Specifically, we prompt the \llm (we use MiniMax-M2.7~\cite{minimaxtwoseven}) with the commit message subject, body, and modified filenames and ask it to classify the operational intent.
To validate our commit intent classifier, one author manually labeled a random sample of 100 commits and compared the human labels against the classifier outputs, yielding a Cohen's kappa~\cite{cohen60} of 0.94.

\mysec{\textnormal{\circled{4}} Lifecycle Analysis}
To study the lifecycle of agentic code at scale, we track all introduced lines and determine whether they{ have been} modified during our observation window. 
For each line added by a commit, we classify it as (1) \textit{survived}: not modified or removed by any future commits; or (2) \textit{terminated}: modified or removed by a future commit. 
For each terminated line, we also identify the exact commit, referred to as the terminal commit, that performed the modification.

At a high level, our lifecycle analysis consists of a diff-tracking pass followed by a \CodeIn{git blame}-based verification pass.
First, for each added line in a commit, we replay subsequent commits in topological order and update each line's current file and line number as later diffs shift the surrounding code. 
When a later diff touches the tracked line's current location, we provisionally mark the line as \textit{terminated} and record the modifying commit as the terminal commit; otherwise, the line remains \textit{survived}.
The diff-tracking pass provides an efficient first approximation of each line's location and possible terminal commit.
Next, we use \CodeIn{git blame} at the end of the observation window to determine whether the original line is still present. 
If the line is still attributed to the origin commit, we mark it as survived. 
For lines that are not found in the final blame output, we verify or recover the earliest terminal commit by checking candidate commits via iterative \CodeIn{git blames}.
To make this tractable at scale, we cache blame results, batch full-file blame calls, and parallelize blame verification across threads.

\mysec{\textnormal{\circled{5}} Static Analysis}
To study changes in code quality metrics and static warnings, we apply static analysis tools to our collected dataset.
We use three widely used static analysis tools:
(1) \textit{\semgrep}~\cite{semgrep}, rule-based analysis tool for detecting security weaknesses in source code;
(2) \textit{\osvscanner}~\cite{osvscanner}, dependency scanner for identifying known dependency vulnerabilities; and
(3) \textit{\sonarqube}~\cite{sonarqube}, software quality analysis tool for tech-debt metrics.
We apply both \semgrep and \osvscanner after each commit in our dataset.
For each commit, we identify all changed files and create two versions of those files: before the commit and after the commit. 
We then run the analysis tools on both versions and obtain two sets of static findings.
To measure the overall repository maintenance metrics, we use 
\sonarqube to analyze weekly snapshots of each repository during our observation window.

\section{RQ1: Does \aicode code require more maintenance post-merge?}
\label{sec:RQ1}
We measure differences in overall maintenance (RQ1a), types of maintenance (RQ1b), and types of corrective maintenance received and introduced defects (RQ1c).

\subsection{Research Methods}

\mysec{Does \aicode code require \textit{more} maintenance? (RQ1a)}
We first ask whether \aicode contributions are maintained at a different overall rate than \humancode contributions.
We model post-merge maintenance as line-level time-to-event data, treating each line introduced by an origin commit as a subject observed from its date of merge until it is either: (1) \textit{terminated} by a later commit, our event-of-interest representing maintenance activity, or (2) \textit{right-censored}, meaning that it does not experience an event before the end of our observation window.

We use survival analysis, a branch of statistical analysis for modeling time-to-event data~\cite{jenkins05}.
Specifically, we use the Kaplan-Meier estimator~\cite{kaplan58}, a standard non-parametric method for estimating survival functions~\cite{cox84}. 
We estimate the \aicode maintenance burden with a Cox proportional hazards model and report the hazard ratio (HR) where HR $>$ 1 indicates a higher termination rate for \aicode lines than for \humancode lines and HR $<$ 1 indicates a lower termination rate.

\mysec{Does \aicode code require \textit{different types} of maintenance? (RQ1b)}
We next ask whether the \textit{composition} of maintenance differs between \aicode and \humancode code.
We first test whether the distribution of maintenance types differs using a Rao-Scott cluster-corrected chi-square test~\cite{rao81}.
We perform this test at the maintenance-action level where each terminal commit that terminates either \aicode or \humancode lines counts as one maintenance action. 
This captures the composition of maintenance work performed on each group of code.
We exclude terminal commits that maintain code from both groups in the same action, because such commits would violate the independence-of-observations assumption.
In total, we excluded 46,320 cross-group actions and retained 157,388 \humancode and 13,729 \aicode{} maintenance actions for analysis.

To estimate whether the \textit{risk} of experiencing each maintenance type differs by authorship, we use Fine-Gray competing-risks regression~\cite{fine99}.
Unlike standard survival analysis, which models a single event-of-interest, our setting allows each subject, \ie line, to be terminated by exactly one of several mutually exclusive maintenance intent types that compete with one another.
Fine-Gray regression addresses this by modeling the subdistribution hazard, \ie the effect of authorship group on the cumulative incidence of a given type while treating the other types as competing events. 
We fit one Fine-Gray model for each maintenance class and operational intent.
We pair each model with its cumulative incidence function (CIF) which estimates the probability that a line has been terminated by that maintenance intent over time.

\mysec{Does \aicode code require different types of corrective maintenance and introduce more defects? (RQ1c)}
Lastly, we focus specifically on differences in corrective maintenance and introduced defects between \aicode and \humancode contributions.
We test whether the distribution of corrective operational intents differs between the two groups using the same chi-square test as RQ1b.
We then estimate corrective maintenance burden using four Cox proportional hazards models: one for overall corrective maintenance and one for each of the three corrective operational intents to compare the corrective termination hazard of \aicode and \humancode lines.

To measure defects directly, we use static analysis to count the static findings each contribution introduces, with \semgrep for source-code security-related findings and \osvscanner for dependency findings.
For each tool, we fit two Poisson generalized linear mixed models (GLMMs)~\cite{mcculloch08} to answer whether \aicode commits introduce more findings per unit of code changed overall, and whether they introduce more high-severity findings.

\mysec{Modeling Considerations}
Because we estimate the \textit{total} effect of authorship, we adjust for confounders in our modeling formulas~\cite{vanderweele19}.
We treat file primary role and repository as confounders: file role reflects the task a contribution serves, while repositories differ in both their level of \aicode adoption and their baseline maintenance behavior.
Repository is a high-cardinality confounder, which we handle using the mechanism appropriate for each model family: stratification by repository in the Cox and Fine-Gray models~\cite{zhou11}, and a repository random intercept in the Poisson GLMMs~\cite{bolker09}.
Within-repository correlation additionally threatens the independence-of-observations assumption. 
We address this by clustering standard errors by repository in the survival models, applying the Rao-Scott correction in the chi-square tests, and including the random intercept in the GLMMs.

The Cox and Fine-Gray models use the same specification: authorship group as the focal covariate, file primary role as a covariate, stratification by repository, and repository-clustered standard errors.
We assess the proportional-hazards assumption for every survival model with the Grambsch-Therneau test and scaled Schoenfeld residual plots~\cite{grambsch94}.
For the overall corrective Cox model, the test flagged a minor proportional-hazards violation.
A time-split robustness check at 21 days found no statistically distinguishable difference between the early and late \aicode effects, so we report the time-averaged hazard ratio as the primary estimate.
For each Fine-Gray model, we verify the minimum events-per-variable threshold of 10 following common statistical best practices~\cite{peduzzi95}, which all categories meet.
Because \aicode lines skew younger, we conduct a calendar-period sensitivity check by re-estimating survival on the overlapping window in which both groups are well represented (August 2025 to May 2026) and comparing the \aicode-\humancode survival gap at 90, 180, and 270 days against the full-window model. 
The gaps match in sign and magnitude, indicating that no calendar adjustment is needed.

The static analysis GLMMs are fit at the commit level with a repository random intercept and a log-exposure offset for the number of relevant changed lines: source-code lines for \semgrep findings and dependency-file lines for \osvscanner findings.
We assess fit with DHARMa simulated residuals and a Pearson overdispersion check~\cite{hartig16}.

\mysec{Limitations}
First, our authorship classification relies on commit-level heuristic signals which may misclassify authorship in either direction.
In addition, human contributions in our dataset may have unobservable AI assistance such as IDE inline suggestions or pasted \aicode outputs.
This makes the \humancode baseline a lower bound on true AI involvement and may bias estimated differences between \aicode and \humancode contributions toward the null.
Second, our maintenance intent labels are obtained from a rule-based and \llm-assisted classification pipeline. 
Particularly, commits near category boundaries (\eg corrective versus perfective) may be mislabeled.
Third, not every line termination necessarily reflects a defect or quality problem; 
we mitigate this by separately analyzing different maintenance intents.
Fourth, our static analysis results are limited to weaknesses and vulnerabilities detectable by the rule-based tools we use and may include false positives.
Finally, our study spans from May 2025 to May 2026 and focuses on active open-source projects with sustained agentic contributions.
As such, longer-horizon maintenance beyond our observation window is unobserved, and our findings may not generalize to closed-source software or projects with minimal \aicode adoption.

\subsection{Results}

\begin{figure}[t]
  \centering
  \begin{subfigure}[t]{0.49\linewidth}
    \includegraphics[width=\linewidth]{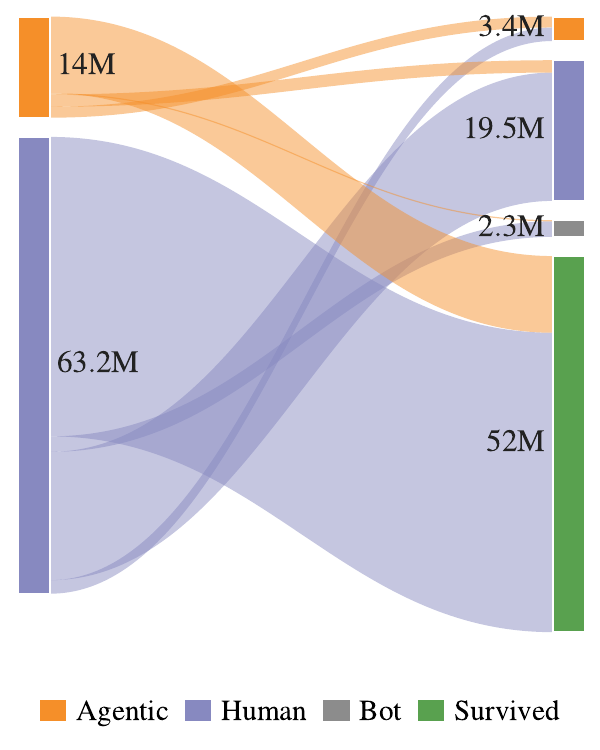}
    \caption{Flow of survival and termination events.}
    \label{fig:rq1a_visualization_sankey}
  \end{subfigure}%
  \begin{subfigure}[t]{0.49\linewidth}
    \includegraphics[width=\linewidth]{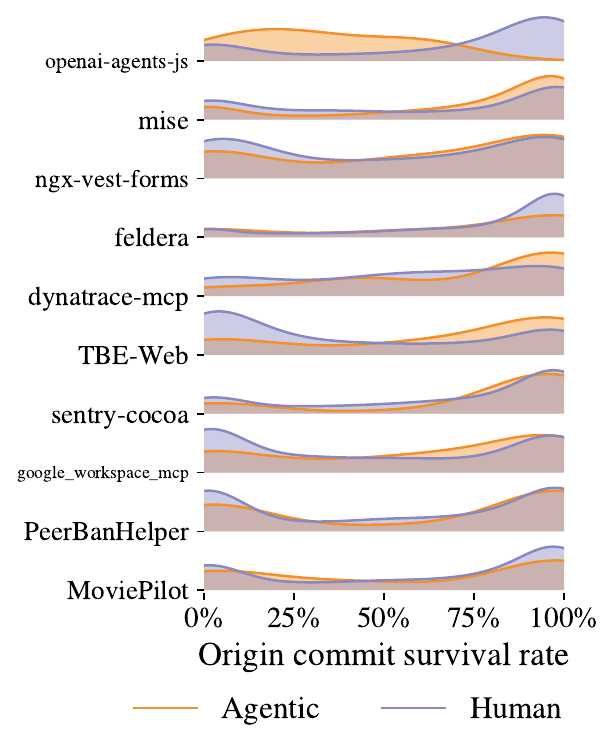}
    \caption{Survival rates of 10 selected repositories.}
    \label{fig:rq1a_visualization_ridgeline}
   \end{subfigure}
  \caption{Post-merge maintenance rate.}
  \label{fig:rq1_visualization}
\end{figure}

\mysec{Does \aicode code require \textit{more} maintenance? (RQ1a)}
\label{mysec:rq1a}
Our results indicate that \aicode code does not carry a uniformly higher overall maintenance burden.
Agentic lines are not significantly more or less likely to be terminated than \humancode lines (HR = 1.11, 95\% CI: 0.85--1.45, $p$ = 0.45).
Although the HR suggests an $\sim$11\% higher termination rate for \aicode lines, the confidence interval includes 1, so the difference is not statistically significant.
This null result is not due to sparse data: in the raw survival rate, \aicode lines survive at least as often as \humancode lines (75.8\% vs. 65.6\%, \cf~Figure~\ref{fig:rq1a_visualization_sankey}).

The flat HR is not evidence that authorship does not matter but rather that its effect is highly project-dependent.
Authorship effects vary widely across per-repository survival distributions (\cf~Figure~\ref{fig:rq1a_visualization_ridgeline}) and this heterogeneity washes out in the aggregate.
We treat this heterogeneity as a key finding in its own right and return to it in more detail in RQ2.

The Sankey flow reveals an asymmetry in who performs terminations: although \humancode commits vastly outnumber \aicode commits, nearly half of \aicode line terminations are performed by \aicode commits themselves (\cf~Figure~\ref{fig:rq1a_visualization_sankey}).
This suggests that agents tend to operate on existing \aicode code and in a different part of the codebase than humans.
Thus, similar overall termination rates do not imply that the underlying maintenance patterns are the same.
We therefore next examine \textit{what kinds} of maintenance each group attracts.

\begin{keyInsightsBox}
    \textbf{Key Insight:} 
     Agentic contributions do not carry a uniform post-merge maintenance burden. 
     Their termination behavior varies substantially across repositories, making project context critical for interpreting post-merge maintenance.
\end{keyInsightsBox}

\begin{figure}
    \centering
    \includegraphics[width=\columnwidth]{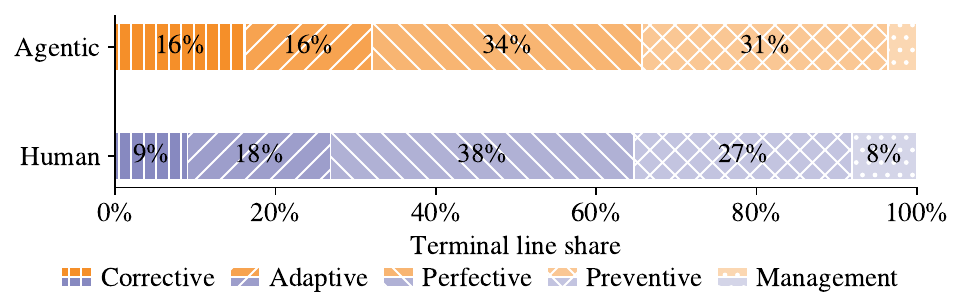}
    \caption{Line-level termination maintenance class proportion.}
    \label{fig:rq1b_visualization_distribution}
\end{figure}

\mysec{Does \aicode code require \textit{different types} of maintenance? (RQ1b)}
\label{mysec:rq1b}
We find that the \textit{composition} of maintenance does differ significantly.
Agentic and \humancode code receive significantly different mixes of maintenance (Rao-Scott chi-square = 910.6, $p <$ 0.001, Cramér's V = 0.07), driven by \aicode code receiving a higher share of bug fixes and a lower share of feature work and refactoring (\cf~Figure~\ref{fig:rq1b_visualization_distribution}).

\begin{figure}
    \centering
    \includegraphics[width=\columnwidth]{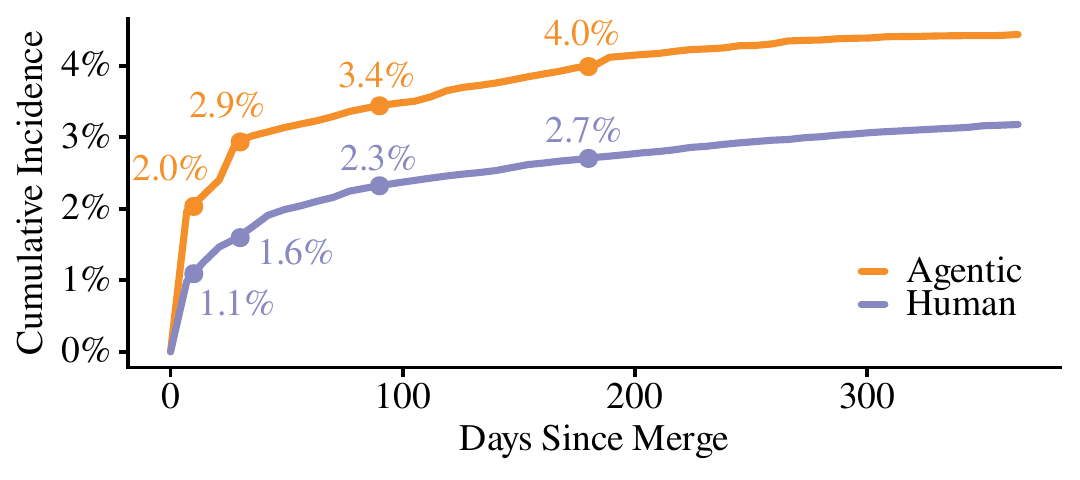}
    \caption{CIF for bug fix. Dots represent time points of 10/30/90/180 days.}
    \label{fig:rq1a_cif_bug_fix}
\end{figure}

The Fine-Gray models confirm this difference appears in \textit{risk} as well: \aicode lines have a significantly higher risk of experiencing corrective maintenance and lower risk of management maintenance (\cf~Table~\ref{tab:rq1a_fine_gray}).
These are the only maintenance classes in which the two groups diverge significantly.
Adaptive, perfective, and preventive maintenance, by contrast, are indistinguishable once file role is accounted for.
At the operational intent level, the elevated corrective risk is specific to bug fixes: \aicode lines reach a higher cumulative incidence of bug-fix termination almost immediately after merge and stay there, with 4.0\% of \aicode lines changed by a bug-fixing commit by 180 days compared with 2.7\% of \humancode lines (\cf~Figure~\ref{fig:rq1a_cif_bug_fix}).
Conversely, \aicode lines are significantly \textit{less} likely to be terminated by style/formatting, merge/release/versioning, and build/config/CI commits.

\begin{table}[t]
  \centering
  \scalebox{1}{
  \begin{tabular}{ll}
\toprule
Maintenance Intent & sHR (95\% CI) \\
\midrule
\textbf{Corrective} & 1.46 [1.02, 2.08]$^{*}$ \\
\hspace{1em}Security fix & 1.18 [0.81, 1.72] \\
\hspace{1em}Revert & 1.39 [0.55, 3.49] \\
\hspace{1em}Bug fix & 1.45 [1.02, 2.07]$^{*}$ \\
\midrule
\textbf{Adaptive} & 1.00 [0.77, 1.31] \\
\hspace{1em}Dependency update & 1.00 [0.77, 1.31] \\
\midrule
\textbf{Perfective} & 0.98 [0.65, 1.50] \\
\hspace{1em}Performance & 1.34 [0.65, 2.80] \\
\hspace{1em}Feature & 0.71 [0.49, 1.02] \\
\hspace{1em}Documentation & 1.17 [0.86, 1.59] \\
\hspace{1em}Resource & 1.81 [0.68, 4.83] \\
\midrule
\textbf{Preventive} & 1.11 [0.69, 1.78] \\
\hspace{1em}Refactor & 0.75 [0.54, 1.04] \\
\hspace{1em}Test & 1.54 [0.96, 2.47] \\
\hspace{1em}Style/Formatting & 0.48 [0.33, 0.68]$^{***}$ \\
\midrule
\textbf{Management} & 0.39 [0.25, 0.60]$^{***}$ \\
\hspace{1em}Merge/Release/Versioning & 0.28 [0.16, 0.51]$^{***}$ \\
\hspace{1em}Build/Config/CI & 0.58 [0.34, 0.99]$^{*}$ \\
\bottomrule
\end{tabular}

  }\par\vspace{0.5em}
  {Note: $^{*} p < 0.05; ^{**}p < 0.01; ^{***}p < 0.001$\par}
  \caption{Fine-Gray regression modeling summary.}
  \label{tab:rq1a_fine_gray}
\end{table}

\begin{keyInsightsBox}
    \textbf{Key Insight:} Agentic and \humancode contributions undergo different types of maintenance, with \aicode code receiving more corrective maintenance.
\end{keyInsightsBox}

\begin{figure}[h]
  \centering
  \begin{subfigure}[t]{0.49\linewidth}
    \includegraphics[width=\linewidth]{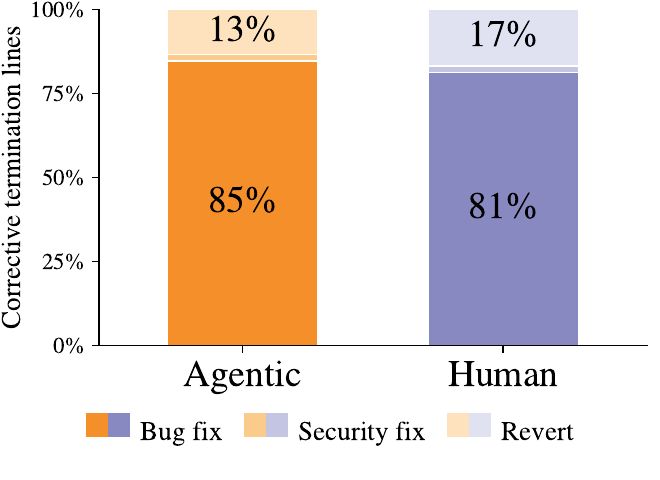}
    \caption{Proportion of the line-level corrective operational intent.}
    \label{fig:rq1c_visualization_corrective_distribution}
  \end{subfigure}%
  \begin{subfigure}[t]{0.49\linewidth}
    \includegraphics[width=\linewidth]{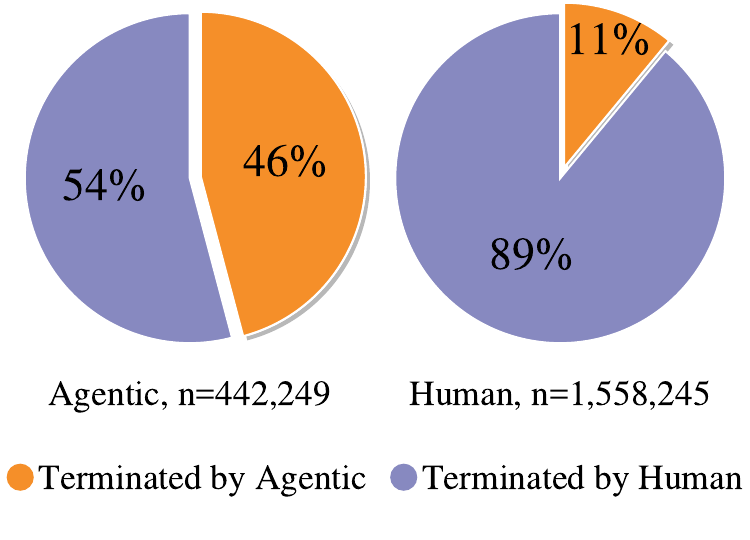}
    \caption{Authorship of bug-fix terminations.}
    \label{fig:rq1c_visualization_bug_fix_terminal_actor}
   \end{subfigure}
  \caption{Post-merge corrective maintenance.}
  \label{fig:rq1c_visualization}
\end{figure}

\begin{table}[t]
  \centering
  \scalebox{0.87}{
  \begin{tabular}{llll}
\toprule
Cox PH & HR (95\% CI) & Poisson GLMM & RR (95\% CI) \\
\midrule
Corrective & 1.49 [1.05, 2.12]$^{*}$ & Static find. & 1.14 [1.08, 1.21]$^{***}$ \\
Bug fix & 1.51 [1.06, 2.16]$^{*}$ & High-sev. static find. & 1.51 [1.33, 1.70]$^{***}$ \\
Security fix & 1.20 [0.82, 1.75] & Dep. find. & 1.10 [1.06, 1.15]$^{***}$ \\
Revert & 1.42 [0.57, 3.54] & High-sev. dep. find. & 1.15 [1.08, 1.23]$^{***}$ \\
\bottomrule
\end{tabular}

  }\par\vspace{0.5em}
  {\raggedleft Note: $^{*} p < 0.05; ^{**}p < 0.01; ^{***}p < 0.001$\par}
  \caption{Cox PH and Poisson GLM models for corrective maintenance and static analysis findings.}
  \label{tab:rq1c_coxph_poisson}
\end{table}

\mysec{Does \aicode code require different types of corrective maintenance and introduce more defects? (RQ1c)}
\label{mysec:rq1c}
We find that \aicode lines receive corrective maintenance at a 49\% higher rate than \humancode lines (\cf~Table~\ref{tab:rq1c_coxph_poisson}).
This gap appears throughout the post-merge period: both groups accumulate fixes fastest in the first few weeks, the \aicode curve rises above the \humancode curve early, and the two stay roughly parallel thereafter (\cf~Figure~\ref{fig:rq1b_cif_corrective}).
This elevated corrective maintenance rate is concentrated in bug fixes. 
Agentic lines show a 51\% higher bug-fix termination rate, while the security-fix and revert effects are directionally higher but not significant (\cf~Table~\ref{tab:rq1c_coxph_poisson}).
The \textit{composition} of corrective work, however, is similar across groups: the bug-fix/security-fix/revert split does not differ significantly (Rao-Scott chi-square = 8.45, $p$ = 0.385, Cramér's V = 0.01; \cf~Figure~\ref{fig:rq1c_visualization_corrective_distribution}). 
Together these findings suggest \aicode code is corrected \textit{more} but not \textit{differently}.

We also find close to half of the bug-fix terminations on \aicode lines come from \aicode commits (\cf~Figure~\ref{fig:rq1c_visualization_bug_fix_terminal_actor}), echoing the RQ1a pattern of agents operating on existing \aicode code. 
Still, more than half of \aicode bug fixes are performed by humans, indicating that \aicode contributions continue to draw substantial human corrective effort.

\begin{figure}[t]
  \centering
  \begin{subfigure}[t]{0.49\linewidth}
    \includegraphics[width=\linewidth]{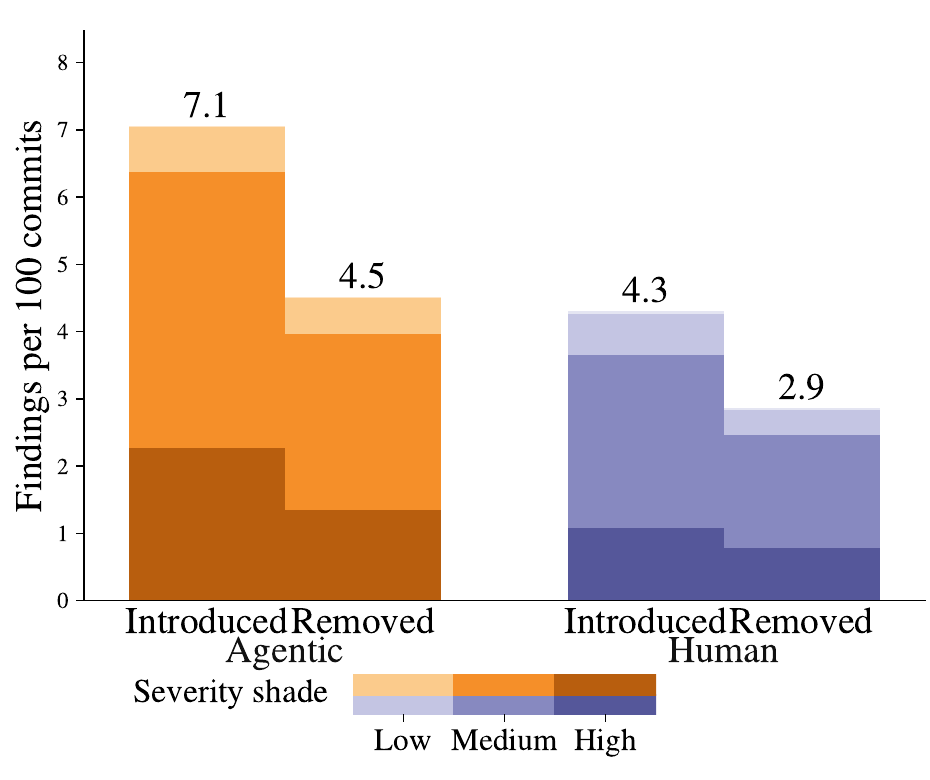}
    \caption{\semgrep findings.}
    \label{fig:rq1c_visualization_semgrep_severity}
  \end{subfigure}%
  \begin{subfigure}[t]{0.49\linewidth}
    \includegraphics[width=\linewidth]{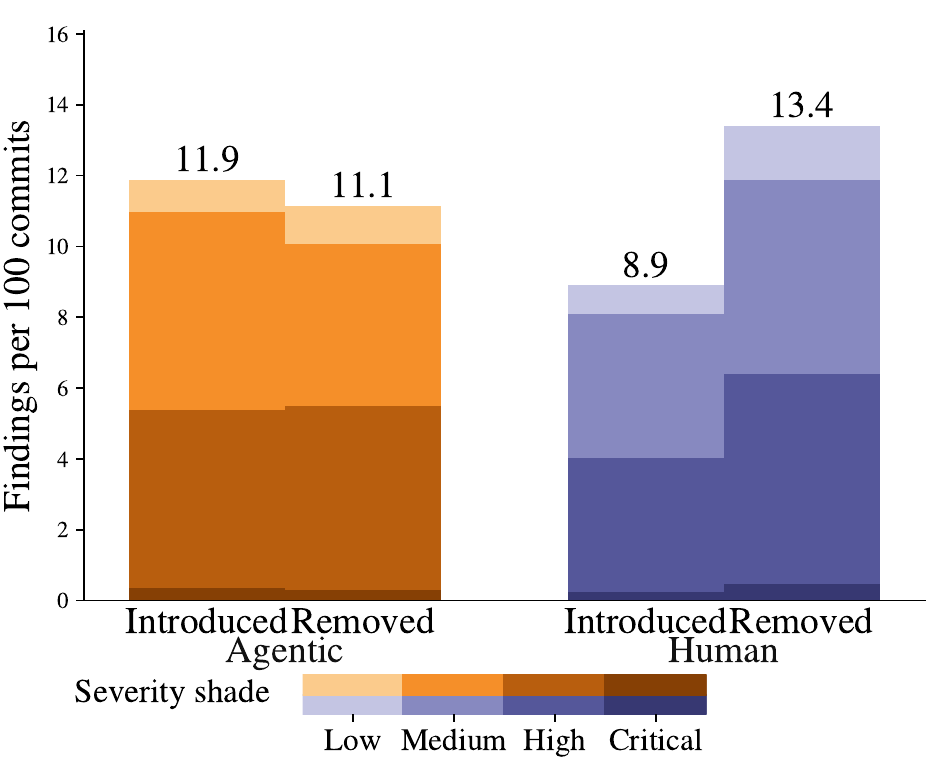}
    \caption{\osvscanner findings.}
    \label{fig:rq1c_visualization_osv_severity}
   \end{subfigure}
  \caption{Static analysis findings per 100 commits.}
  \label{fig:rq1c_visualization_static}
\end{figure}

Additionally, we examine the source-code and dependency findings introduced by \aicode and \humancode code.
Agentic contributions introduce \semgrep security-related findings at 1.14 times and high-severity findings (\CodeIn{ERROR}) at 1.51 times the \humancode per-source-line rate (\cf~Figure~\ref{fig:rq1c_visualization_semgrep_severity} and Table~\ref{tab:rq1c_coxph_poisson}).
The same holds for dependency vulnerabilities: \aicode contributions introduce \osvscanner findings at 1.10 times and high-severity findings (\CodeIn{HIGH}/\CodeIn{CRITICAL}) at 1.15 times the \humancode per-dependency-line rate (\cf~Figure~\ref{fig:rq1c_visualization_osv_severity} and Table~\ref{tab:rq1c_coxph_poisson}).

\begin{keyInsightsBox}
    \textbf{Key Insight:} Agentic code faces higher corrective-maintenance risk, driven primarily by bug fixes. 
    Agentic commits also introduce weaknesses and vulnerabilities at higher rates, especially high-severity findings.
\end{keyInsightsBox}

\section{RQ2: Does corrective maintenance burden increase in projects as the proportion of agentic code increases over time?}
\label{sec:RQ2}
We now examine whether corrective maintenance accumulates as \aicode code share increases.
To do so, we conduct a longitudinal panel study that analyzes how increases in \aicode code share affect repository-level corrective maintenance.

\subsection{Research Methods}

We ask whether repositories experience more corrective maintenance as \aicode code accounts for a larger share of their codebase over time.
We construct a longitudinal panel dataset that follows repositories month by month, allowing us to relate changes in \aicode code share to subsequent corrective maintenance activity.
We use repository-month as the unit of analysis.
For each repository $i$ and month $m$, we measure corrective maintenance using two outcomes: 
(1) $\textit{corrective\_commit\_rate}_{i,m}$: number of corrective commits / total number of commits and
(2) $\textit{human\_corrective\_commit\_rate}_{i,m}$: the number of human corrective commits / total number of human commits.
At the same time, we compute the $\textit{agentic\_code\_share}_{i,m}$ as the percentage of the repository's code, measure at the end of the month, that was written by agents. 
Using these monthly measures, we fit panel models based on Binomial generalized linear mixed models (GLMMs)~\cite{mcculloch08} and report odds ratios (ORs).
In our setting, OR $>$ 1 indicates that the predictor increases corrective maintenance rate, while OR $<$ 1 means predictor decreases corrective maintenance rate.
Our key predictor is lagged \aicode code share ($\textit{agentic\_code\_share}_{i,m-1}$) which captures the \aicode code share at the end of the previous month.
We use lagged \aicode code share because our goal is to measure how prior \aicode code presence is associated with corrective maintenance activity in the following month.

\mysec{Modeling Considerations}
For the Binomial GLMMs used in our panel analysis, we first control for commit volume using the log-transformed number of commits: total and \humancode commits for the overall and \humancode corrective commit rate models respectively.
We include a random intercept for each repository, allowing each repository to have its own baseline corrective commit rate.
We also include an observation-level random effect (OLRE)~\cite{harrison14} to account for overdispersion in the binomial response, absorb extra-binomial variation and reduce the risk of understated standard errors.
Additionally, we performed model diagnostics to assess fit and robustness, including singular-fit checks, residual diagnostics, Pearson overdispersion tests, and grouped-binomial zero-success checks comparing the observed number of zero-corrective repository-months against the fitted model expectation.

\mysec{Limitations}
Our model uses one-month lag which captures only a coarse temporal relationship where we do not model corrective maintenance that may respond to \aicode code over shorter or longer windows.
Additionally, our metric of corrective commit rate only measures one aspect of corrective burden but not the size, complexity, or review effort to perform those fixes. 
As such, our findings should be generalized cautiously.

\subsection{Results}

\begin{figure}[t]
  \centering
  \includegraphics[width=\columnwidth]{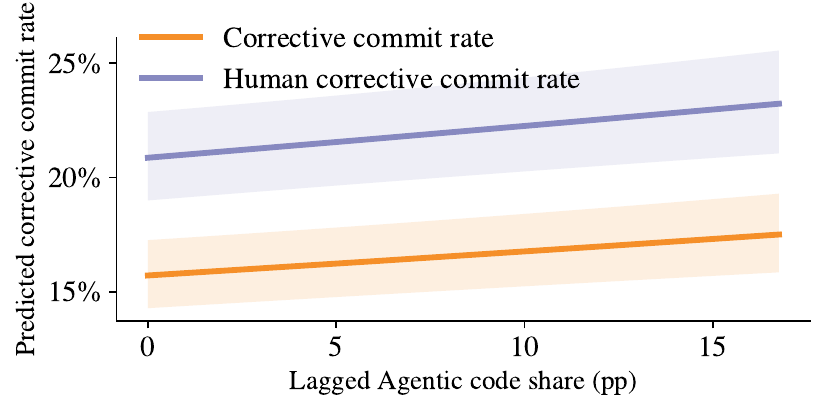}
  \caption{Predicted corrective maintenance rates as a function of lagged \aicode code share. 
  Slopes show Binomial GLMM's predicted rates for all corrective commits and \humancode corrective commits across repo-months, with shaded bands indicating 95\% confidence intervals.}
  \label{fig:rq3_visualization}
\end{figure}

Both the overall corrective commit rate and the human corrective commit rate increase as lagged agentic code share rises (\cf~Figure~\ref{fig:rq3_visualization}).
Repositories with more existing \aicode code are more likely to spend the following month on both overall corrective maintenance (OR = 1.08, 95\% CI: 1.05--1.12, $p<0.001$) and human corrective maintenance (OR = 1.09, 95\% CI: 1.05--1.13, $p<0.001$).
A 10 percentage-point increase in lagged \aicode code share is associated with $\sim$8.0\% and $\sim$8.6\% higher odds of corrective commit overall and human corrective commits in the following month respectively.
This indicates greater \aicode code presence is associated not only with higher repository-level corrective burden but also with higher corrective burden for human developers.

\begin{keyInsightsBox}
    \textbf{Key Insight:} Corrective maintenance burden increases with \aicode code share, including corrective maintenance performed by human developers.
\end{keyInsightsBox}

\section{RQ3: What project characteristics are associated with agentic maintenance burden?}
\label{sec:RQ3}

RQ1 shows that the maintenance burden of \aicode code varies substantially across projects.
This heterogeneity suggests that project characteristics may affect the \aicode maintenance burden.
In this research question, we concretely examine which project characteristics explain differences in post-merge maintenance burden between \aicode and \humancode contributions.

\subsection{Research Methods}

To examine which project characteristics explain variation in \aicode maintenance burden, we use a two-stage meta-regression approach.
The key idea is to first estimate a separate \aicode maintenance burden for each project and then test whether project characteristics explain why this burden is larger in some projects than in others.
In stage one, for each project in our dataset, we fit a Cox PH model to measure the \aicode maintenance burden relative to \humancode lines, expressed as a hazard ratio (HR).
This follows RQ1a, but instead of estimating a single HR for the entire dataset, we compute an HR and Standard Error (SE) for each project. 
In stage two, we use each project's HR as the outcome weighted by their uncertainty (SE) to model whether the different project characteristics explain variation in \aicode maintenance burden using a random-effects meta-regression model.
We operationalize six project characteristics, each corresponding to a hypothesis about factors that may affect \aicode maintenance burden:

\begin{itemize}[leftmargin=*, label={}]
    \item \textbf{Technical Debt:} 
    We use the technical debt index computed by \sonarqube at the start of the observation window. 
    This is a proxy for the project's baseline maintenance practices. 
    We expect projects with high technical debt indexes to have a harder time integrating new \aicode code cleanly. 
    \item \textbf{Number of Contributors:}
    We count the number of non-bot contributors with at least one commit during the observation window.
    We expect projects with more contributors to have lower \aicode maintenance burden as more maintainers may have more review bandwidth to catch issues pre-merge.
    \item \textbf{Agentic Source Code Share:}
    We measure the fraction of \aicode churn that occurs in files classified as source code.
    We expect projects with higher \aicode source-code share to have higher \aicode maintenance burden, because \aicode activity concentrated in source code is more directly exposed to post-merge maintenance than activities in lower-stake documentation, configuration, or resource changes.
    \item \textbf{Test File Change Share:}
    We measure the share of changed files classified as tests.
    We expect projects with higher test file change share to have lower \aicode maintenance burden since projects with stronger test infrastructure can catch and stabilize \aicode code faster.
    \item \textbf{Agentic Line Churn Share:}
    We measure the fraction of total line churn in the project by \aicode contributions.
    We expect projects with higher \aicode line-churn share to have higher \aicode maintenance burden, because heavier \aicode adoption can become load-bearing in the codebase and therefore more likely to be revisited and modified.
    \item \textbf{No Review Rate:}
    We measure the proportion of merged pull requests during the observation window that received no code review.
    We expect projects with higher no-review rate to have higher \aicode maintenance burden, because weaker review processes are less able to catch problematic or low quality \aicode contributions (i.e., AI slop).
\end{itemize}

\mysec{Modeling Considerations}
For stage one, we fit the Cox PH models similar to RQ1a.
However, we do not stratify or cluster based on repository because we compute a separate HR for each repository.
We then filter out projects with fewer than 100 tracked lines or 10 termination events in either groups (5 projects).
This avoids unstable project-level HR estimates driven by sparse data.
For the stage two random-effects meta-regression model, we use the Knapp-Hartung correction~\cite{hartung01} following standard best practices.
To determine which covariates to operationalize, we conduct several exploratory analyses, including data visualization and LASSO-based screening~\cite{tibshirani96} of candidate covariates.
Based on these analyses, we select the six covariates described above.
We perform standard model diagnostics, including checks for multicollinearity, influential projects, and residual normality.

\mysec{Limitations} 
Our covariates are necessarily incomplete and proxy-based: some relevant project characteristics, such as review quality, are not directly observable in our dataset. Accordingly, the meta-regression should be interpreted as identifying associations rather than fully explaining cross-project variation in agentic maintenance burden.

\subsection{Results}

\begin{table}[t]
  \centering
  \scalebox{1}{
  \begin{tabular}{ll}
\toprule
Covariate & HR multiplier (95\% CI) \\
\midrule
Technical Debt & 1.47 [0.92, 2.36] \\
Number of Contributors & 1.00 [1.00, 1.00] \\
Agentic Source Code Share & 0.72 [0.46, 1.13] \\
Test File Change Share & 1.90 [0.85, 4.26] \\
Agentic Line Churn Share & 0.73 [0.40, 1.34] \\
No Review Rate & 1.75 [1.01, 3.04]$^{*}$ \\
\bottomrule
\end{tabular}

  }\par\vspace{0.5em}
  {Note: $^{*} p < 0.05; ^{**}p < 0.01; ^{***}p < 0.001$\par}
  \caption{Random-effect meta-regression model with $I^2$ = 100\%, $R^2$ = 6.0\%. 
          }
  \label{tab:rq3_meta_regression}
\end{table}

Across 177 projects, we find statistically significant evidence that \aicode maintenance burden is associated with project characteristics rather than varying at random ($F$(6, 170) = 2.86, $p <$ 0.05).
The rate of merging code without review emerges as the strongest individual predictor (\cf~Table~\ref{tab:rq3_meta_regression}), with higher no-review rates being associated with significantly higher \aicode maintenance burden. 
In particular, each 10 percentage-point increase in a project's no-review rate is associated with roughly a 6\% increase in agentic maintenance burden, on average. 
Technical debt and limited test infrastructure show non-significant effects in the same direction (\cf~Table~\ref{tab:rq3_meta_regression}), suggesting a coherent pattern in which weaker engineering safeguards are associated with increased \aicode maintenance burden.

\begin{keyInsightsBox}
    \textbf{Key Insight:}
    Agentic maintenance burden is associated with project characteristics and particularly, 
    having higher no-review rate is associated with increased \aicode maintenance burden.
\end{keyInsightsBox}

\section{Discussion and Implications}
\label{sec:discussion}

\mysec{The Real Asymmetry}
The adoption of \aicode coding tools shattered a core symmetry between code generation and review that software development long relied on: code generation and code review were both constrained by human labor.
Before GenAI, developers could only produce code as fast as they could understand, write, and revise it; reviewers, in turn, could only approve code as fast as they could understand its behavior and consequences.
The entire pitch of \aicode coding tools is that they decouple generation output velocity from human labor: a person can now produce far more code per unit of human effort than they could have by hand. 
But review did not get the same decoupling.
The result is a generation-review velocity asymmetry that exists by construction: generation has been severed from the human constraint that used to pace it, while review remains bound to human comprehension.

This asymmetry is starkly visible in practice today.
As teams adopt agentic tools, pull request volume increases while review time rises and review thoroughness and rates decline~\cite{faros26}. 
Our results help show the cost of this asymmetry.

\mysec{Why a Full-Lifecycle View is Needed to See It}
Although we do not measure the generation-review asymmetry directly, we do measure its consequences.
Observing the consequences requires following \aicode code throughout its entire lifecycle rather than stopping at merge~\cite{watanabe25, peralta26, nakashima26, haider26, minh26, watanabe26}, or only looking after merge~\cite{liu26, rahman26}.
Initially, \aicode code looks reassuring with a similar overall maintenance rate to human code~(Section~\ref{mysec:rq1a}).
That similarity does not survive a closer look.
Agentic code receives more corrective maintenance with more bug fixes~(Section~\ref{mysec:rq1b}), introduces more security weaknesses and dependency vulnerabilities~(Section~\ref{mysec:rq1c}), and is associated with higher overall corrective burden in projects as its share grows~(Section~\ref{sec:RQ2}). 
Additionally, projects that merge \aicode code without review show the largest \aicode maintenance burden of all~(Section~\ref{sec:RQ3}).

Because these consequences are only visible when tracing \aicode code across its full lifecycle, precisely where most current evaluation does not look, adoption-informing metrics must expand beyond generation velocity and pre-merge performance.
Today, success is often measured through visible and easy-to-count signals: pull request volume, code generated, percentage of \aicode code, and token consumption.
But our results show that those measures are blind to the outcome that matters most for software development: whether the code is reliable, secure, and maintainable after merge.

\mysec{The Review Side Cannot Close the Gap}
Faced with the generation-review asymmetry, the response across the software industry has been to treat it as a review-side problem and to adjust review processes until they keep pace with how fast code is produced by \aicode tools~\cite{roychoudhary25}. 
These responses cluster into two broad strategies and neither closes the gap on its own.

The first is to scale human review, which runs into a limit at every point on its range no matter how much you attempt to scale it, because a human remains the final checkpoint. 
At the extreme, projects that merge \aicode code without review at high rates are precisely the projects where \aicode maintenance burden is the highest~(Section~\ref{sec:RQ3}). 
Short of that extreme, lighter and faster review removes fewer defects~\cite{mcintosh16}, and \aicode code introduces more defects to begin with~(Section~\ref{sec:RQ1}).

The second option is to automate review.
Automation can resolve the asymmetry, but the asymmetry is created exactly because review is the last checkpoint where a human decides whether an agents output is acceptable, and automating it surrenders that decision making autonomy to the same kind of system that produced the code rather than restoring human control of the generation velocity.
We do not claim that automated review is inherently worse than human review; our data does not measure that. 
This point is instead structural: review is currently the only thing that keeps the rate of what enters a project tied, however loosely, to a rate a person can keep up with.
Automating it removes all human control over development velocity and once a project depends on running at the automated rate, restoring a human-based checkpoint means falling behind on a volume of output the project has already organized itself around, which makes it an extremely difficult decision to walk back in practice.

Even if one is willing to cede human decision making autonomy, GenAI automated review systems do not magically solve the problem. 
These review systems are themselves software that must be evaluated and secured. 
Recently, the popular Claude Code GitHub action used for automated pull request review has been shown to expose CI/CD secrets after encountering untrusted content~\cite{microsoft26}.
In other words, building automated review tools does not alleviate the problem, but instead can introduce additional development burdens.

\mysec{Fix the Tool, Not the Process}
At this point our argument moves from what we measured to what we believe the mechanism implies. 
If agentic tools increase generation velocity while human review continues to bound review velocity, 
then the two sides of the development pipeline are no longer paced by the same constraint. That mismatch cannot be absorbed by review indefinitely.
This is why a hybrid configuration that keeps humans in control of review while GenAI controls generation is not sustainable. 
Eventually, the two process velocities must be brought under the same control, or the system breaks down under its own backlog. 
Adding GenAI to the review side may help with triage or prioritization, but it does not by itself resolve the deeper asymmetry if humans remain accountable for final judgment. 
Instead, it defers the cost while projects build a growing dependence on increasingly automated generation processes.
We cannot come to a conclusion about the impacts of this tension with our data, but we raise it as a central question for longitudinal work on \aicode tooling adoption.

If the asymmetry cannot be resolved until the entities controlling the output velocity of both processes are matched, and we continue to build a dependence on GenAI on the generation side, what does that mean for the moment when this asymmetry comes to a head and we are forced to make a decision?
This question is not abstract.
In a growing number of settings, declining to adopt is no longer a realistic option. 
Shopify and Duolingo have made \agentic code use a baseline expectation~\cite{palmer25,shibu25}, Meta has set internal GenAI-use targets which include goals for \aicode code changes and broad adoption among engineers~\cite{langley26b}, and developers at companies such as Microsoft report being evaluated in part by how much \aicode tooling they use~\cite{stewart25}.
Where opting out is unavailable, teams cannot avoid the asymmetry; they can only manage it.
If this mechanism holds, the choice is narrowed in advance: a team that leans on generation to stay ahead of review builds a dependence that makes the tool harder to question later, at the exact moment when such questioning and critical reflection would matter most. 

The risk itself lies in the misconception that a stable hybrid approach exists without sufficient evidence while rapidly developing a dependence on the technology. 
If we as a collective are serious about preserving human control over digital infrastructure and decision making autonomy, then generation velocity must remain governable by humans.
The tools themselves must be designed to respect the limits of human comprehension, validation, and accountability. 
We must fix the tools, not merely stretch the process around them.

\subsection{Implications for Research and Practice}
For researchers, our results argue that lifecycle-spanning evaluation of \aicode code should become a standard complement to pre-merge benchmarks.
Merge-time success is not enough: evaluations should also measure whether agentic contributions remain durable, secure, and maintainable.
This aligns with recent practitioner calls for standardized first-class outcome metrics of post-merge maintenance durability, \eg \textit{code durability}, and \textit{code turnover rate}.

For practitioners, our results argue for protecting review rather than treating it as {something} to compress.
Recognize that review alone, so long as it remains a process humans have ultimate decision making autonomy over, cannot indefinitely absorb a source no longer bound by human effort.

For organizations making GenAI adoption decisions, our results urge caution and reflection on what metrics are being used to define the success and to ensure their visibility is in alignment with the metrics of success for the organization's software development practices. 

For tool builders, our results argue that the decisive measure of a coding agent is not whether its output merges, but what {it} costs the project over time.
The intervention is therefore to reduce that cost at the source: tools should be designed not just to generate more code, but to generate code that survives, imposes less maintenance burden, and remains sustainable post-merge.

\section{Conclusion}
\label{sec:conclusion}

In this work, we conducted a longitudinal post-merge study of \aicode and \humancode contributions across 182 repositories using line-level survival, maintenance intent, static analysis findings, and project characteristics.
Our results show that while \aicode code does not exhibit uniformly higher overall maintenance, it receives more corrective maintenance, and introduces more security weaknesses and dependency vulnerabilities than human contributions. 
At the project level, increasing \aicode code share is associated with higher corrective maintenance rates and projects that merge more code without review show larger \aicode maintenance burdens.
As generation becomes cheaper, the relevant question is no longer whether agents can write mergeable code, but whether their contributions remain reliable, secure, and maintainable after merge.

\section{Data Availability}
\label{sec:dataavailability}

We provide our complete artifact for reproduction:\break
\url{https://github.com/post-merge-reality/post-merge-reality}

\bibliographystyle{IEEEtran}
\bibliography{cmbib}

@String{ICSE              = "Proc.\ Int'l Conf.\ Software Engineering (ICSE)"}

@String{TOOLS             = "Proc.\ Int'l Conf.\ Objects, Models, Components, Patterns (TOOLS EUROPE)"}

@String{CHI             = "Proc.\ Conf.\ Human Factors in Computing Systems (CHI)"}

@String{MODELS          = "Proc.\ Int'l Conf.\ Model Driven Engineering Languages and Systems (MoDELS)"}

@String{ISSRE =     "Proc.\ Int'l Symp.\ Software Reliability Engineering (ISSRE)"}

@String{MSR =       "Proc. Conf.\ Mining Software Repositories (MSR)"}

@String{EMSE = "Empirical Software Engineering"}

@String{SCAM        = "Proc.\ Int'l Working Conf.\ Source Code Analysis and Manipulation (SCAM)"}

@String{ACM = "ACM Press"}

@String{Springer="Springer-Verlag"}

@String{IEEE="IEEE Computer Society"}

@misc{semgrep, 
 author={Semgrep},
 title = {Semgrep App Security Platform},
 howpublished = {\url{https://semgrep.dev/}},
 note = {Accessed Jun.\ 2026}
}

@misc{osvscanner, 
 author={OSV-Scanner},
 title = {OSV-Scanner},
 howpublished = {\url{https://google.github.io/osv-scanner/}},
 note = {Accessed Jun.\ 2026}
}

@misc{sonarqube, 
 author={SonarQube},
 title = {SonarQube: Fight AI Slop \& Verify AI Code $\vert$ Sonar},
 howpublished = {\url{https://www.sonarsource.com/products/sonarqube/}},
 note = {Accessed Jun.\ 2026}
}

@misc{cursor, 
 author={Cursor},
 title = {Cursor: AI coding agent},
 howpublished = {\url{https://cursor.com/}},
 note = {Accessed Jun.\ 2026}
}

@misc{copilot, 
 author={GitHub},
 title = {GitHub Copilot Your AI pair programmer},
 howpublished = {\url{https://github.com/features/copilot}},
 note = {Accessed Jun.\ 2026}
}

@misc{claudecode, 
 author={Anthropic},
 title = {Claude Code by Anthropic $\vert$ AI Coding Agent, Terminal, IDE},
 howpublished = {\url{https://claude.com/product/claude-code}},
 note = {Accessed Jun.\ 2026}
}

@misc{codex, 
 author={OpenAI},
 title = {Codex $\vert$ AI Coding Partner from OpenAI $\vert$ OpenAI},
 howpublished = {\url{https://openai.com/codex/}},
 note = {Accessed Jun.\ 2026}
}

@misc{minimaxtwoseven,
  title = {{MiniMax M2.7}},
  howpublished = {\url{https://www.minimax.io/models/text/m27/}},
  note = {Accessed Jun.\ 2026}
}

@online{faros26,
  author       = {Research, Faros},
  title        = {Ten takeaways from the AI Engineering Report 2026: The Acceleration Whiplash},
  organization = {Faros},
  year         = {2026},
  url          = {https://www.faros.ai/blog/ai-acceleration-whiplash-takeaways},
}

@article{agarwal26,
  title={AI IDEs or Autonomous Agents? Measuring the Impact of Coding Agents on Software Development},
  author={Agarwal, Shyam and He, Hao and Vasilescu, Bogdan},
  journal={arXiv preprint arXiv:2601.13597},
  year={2026}
}

@article{baltes26,
  title={``An Endless Stream of AI Slop": The Growing Burden of AI-Assisted Software Development},
  author={Baltes, Sebastian and Cheong, Marc and Treude, Christoph},
  journal={arXiv preprint arXiv:2603.27249},
  year={2026}
}

@article{borg26,
  title={Echoes of AI: Investigating the downstream effects of AI assistants on software maintainability},
  author={Borg, Markus and Hewett, Dave and Hagatulah, Nadim and Couderc, Noric and S{\"o}derberg, Emma and Graham, Donald and Kini, Uttam and Farley, Dave},
  journal=EMSE,
  volume={31},
  number={6},
  pages={161},
  year={2026},
  publisher={Springer}
}

@article{bort26,
  author = {Julie Bort},
  title = {Anthropic’s Claude popularity with paying consumers is skyrocketing},
  journal={TechCrunch},
  year = {2026},
  note = "{\url{https://techcrunch.com/2026/03/28/anthropics-claude-popularity-with-paying-consumers-is-skyrocketing/}}"
}

@article{cui26,
  title={The effects of generative AI on high-skilled work: Evidence from three field experiments with software developers},
  author={Cui, Kevin Zheyuan and Demirer, Mert and Jaffe, Sonia and Musolff, Leon and Peng, Sida and Salz, Tobias},
  journal={Management Science},
  year={2026},
  publisher={INFORMS}
}

@article{cynthia26,
  title={Beyond Bug Fixes: An Empirical Investigation of Post-Merge Code Quality Issues in Agent-Generated Pull Requests},
  author={Cynthia, Shamse Tasnim and Muttakin, Al and Roy, Banani},
  journal={arXiv preprint arXiv:2601.20109},
  year={2026}
}

@online{gitlab26,
  author       = {{GitLab Inc.}},
  title        = {{GitLab} Research Reveals Organizations Are Generating {AI} Code Faster Than They Can Control It},
  organization = {GitLab},
  year         = {2026},
  url          = {https://ir.gitlab.com/news/news-details/2026/GitLab-Research-Reveals-Organizations-Are-Generating-AI-Code-Faster-Than-They-Can-Control-It/default.aspx},
}

@online{glover26,
  author       = {Glover, Ellen},
  title        = {What is Tokenmaxxing? The {AI} Workplace Trend Explained},
  organization = {Built In},
  year         = {2026},
  url          = {https://builtin.com/articles/ai-tokenmaxxing},
}

@article{haider26,
  title={Understanding Dominant Themes in Reviewing Agentic AI-authored Code},
  author={Haider, Md Asif and Zimmermann, Thomas},
  journal={arXiv preprint arXiv:2601.19287},
  year={2026}
}

@article{haque26,
  title={Do Autonomous Agents Contribute Test Code? A Study of Tests in Agentic Pull Requests},
  author={Haque, Sabrina and Ingale, Sarvesh and Csallner, Christoph},
  journal={arXiv preprint arXiv:2601.03556},
  year={2026}
}

@inproceedings{he26,
  title={Speed at the Cost of Quality: How Cursor AI Increases Short-Term Velocity and Long-Term Complexity in Open-Source Projects},
  author={He, Hao and Miller, Courtney and Agarwal, Shyam and K{\"a}stner, Christian and Vasilescu, Bogdan},
  booktitle=MSR,
  year={2026}
}

@article{kharma26,
  title={Security and quality in llm-generated code: A multi-language, multi-model analysis},
  author={Kharma, Mohammed F and Choi, Soohyeon and Alkhanafseh, Mohammad and Mohaisen, David},
  journal={IEEE Transactions on Dependable and Secure Computing},
  year={2026},
  publisher={IEEE}
}

@online{langley26,
  author       = {Langley, Hugh},
  title        = {{Google} says 75\% of the company's new code is {AI}-generated},
  organization = {Business Insider},
  year         = {2026},
  url          = {https://www.businessinsider.com/google-ai-generated-code-75-gemini-agents-software-2026-4},
}

@online{langley26b,
  author       = {Langley, Hugh},
  title        = {Meta's {AI} Push Ties Employee Goals to {AI} Tool Adoption},
  organization = {Business Insider},
  year         = {2026},
  url          = {https://www.businessinsider.com/meta-ai-push-employee-goals-tool-adoption-2-026-3},
}

@article{liu26,
  title={Debt behind the ai boom: A large-scale empirical study of ai-generated code in the wild},
  author={Liu, Yue and Widyasari, Ratnadira and Zhao, Yanjie and Irsan, Ivana Clairine and Chen, Junkai and Lo, David},
  journal={arXiv preprint arXiv:2603.28592},
  year={2026}
}

@online{microsoft26,
  author       = {{Microsoft Defender Security Research Team} and Edry, Dor and Eliahu, Amit},
  title        = {Securing {CI/CD} in an agentic world: {Claude Code} {GitHub} action case},
  organization = {Microsoft Security Blog},
  year         = {2026},
  url          = {https://www.microsoft.com/en-us/security/blog/2026/06/05/securing-ci-cd-in-agentic-world-claude-code-github-action-case/},
  urldate      = {2026-07-01}
}

@article{ogenrwot26,
  title={Patchtrack: A comprehensive analysis of chatgpt’s influence on pull request outcomes},
  author={Ogenrwot, Daniel and Businge, John},
  journal=EMSE,
  volume={31},
  number={5},
  pages={136},
  year={2026},
  publisher={Springer}
}

@misc{openai26, 
 author={OpenAI},
 title = {Codex for every role, tool, and workflow},
 howpublished = {\url{https://openai.com/index/codex-for-every-role-tool-workflow/}},
 year={2026}
}

@article{rahman26,
  title={Will It Survive? Deciphering the Fate of AI-Generated Code in Open Source},
  author={Rahman, Musfiqur and Shihab, Emad},
  journal={arXiv preprint arXiv:2601.16809},
  year={2026}
}

@inproceedings{miller26b,
  title={``Maybe We Need Some More Examples:'' Individual and Team Drivers of Developer GenAI Tool Use},
  author={Miller, Courtney and Choudhuri, Rudrajit and Ulloa, Mara and Haniyur, Sankeerti and DeLine, Robert and Storey, Margaret-Anne and Murphy-Hill, Emerson and Bird, Christian and Butler, Jenna L},
  booktitle={Proc.\ Int'l Conf.\ Software Engineering (ICSE)},
year={2026. ACM Distinguished Paper Award},
  shorthand = {ICSE'26B}
}

@article{minh26,
  title={Early-Stage Prediction of Review Effort in AI-Generated Pull Requests},
  author={Minh, Dao Sy Duy and Kiet, Huynh Trung and Quy, Nguyen Lam Phu and Hoa, Pham Phu and Nguyen, Tran Chi and Duong, Nguyen Dinh Ha and Tran, Truong Bao},
  journal={arXiv preprint arXiv:2601.00753},
  year={2026}
}

@article{nakashima26,
  title={Why Agentic-PRs Get Rejected: A Comparative Study of Coding Agents},
  author={Nakashima, Sota and Ishimoto, Yuta and Kondo, Masanari and Mclntosh, Shane and Kamei, Yasutaka},
  journal={arXiv preprint arXiv:2602.04226},
  year={2026}
}

@article{peralta26,
  title={Why Are Agentic Pull Requests Merged or Rejected? An Empirical Study},
  author={Peralta, Sien Reeve O and Hoshi, Fumika and Washizaki, Hironori and Ubayashi, Naoyasu and Kondo, Inase and Higo, Yoshiki and Mukai, Hiroki and Yoshida, Norihiro and Kusama, Kazuki and Tanaka, Hidetake and others},
  journal={arXiv preprint arXiv:2605.22534},
  year={2026}
}

@article{watanabe26,
  title={How AI Coding Agents Communicate: A Study of Pull Request Description Characteristics and Human Review Responses},
  author={Watanabe, Kan and Tsuchida, Rikuto and Monno, Takahiro and Huang, Bin and Yamasaki, Kazuma and Fan, Youmei and Shimari, Kazumasa and Matsumoto, Kenichi},
  journal={arXiv preprint arXiv:2602.17084},
  year={2026}
}

@online{young26,
  author       = {Young, Jim},
  title        = {Introducing the State of {AI} Coding 2026},
  organization = {New Relic},
  year         = {2026},
  url          = {https://newrelic.com/blog/ai/state-of-ai-coding-2026},
}

@article{becker25,
  title={Measuring the impact of early-2025 AI on experienced open-source developer productivity},
  author={Becker, Joel and Rush, Nate and Barnes, Elizabeth and Rein, David},
  journal={arXiv preprint arXiv:2507.09089},
  year={2025}
}

@article{borek25,
  title={Quality evaluation of Tabby coding assistant using real source code snippets},
  author={Borek, Marta and Nowak, Robert},
  journal={arXiv preprint arXiv:2504.08650},
  year={2025}
}

@techreport{github25b,
  author={GitHub},
  title={Octoverse: A new developer joins GitHub every second as AI leads TypeScript to \#1},
  year=2025,
  url={https://github.blog/news-insights/octoverse/octoverse-a-new-developer-joins-github-every-second-as-ai-leads-typescript-to-1/},
  institution={GitHub}
}

@article{horikawa25,
  title={Agentic Refactoring: An Empirical Study of AI Coding Agents},
  author={Horikawa, Kosei and Li, Hao and Kashiwa, Yutaro and Adams, Bram and Iida, Hajimu and Hassan, Ahmed E},
  journal={arXiv preprint arXiv:2511.04824},
  year={2025}
}

@article{li25,
  title={The rise of ai teammates in software engineering (se) 3.0: How autonomous coding agents are reshaping software engineering},
  author={Li, Hao and Zhang, Haoxiang and Hassan, Ahmed E},
  journal={arXiv preprint arXiv:2507.15003},
  year={2025}
}

@online{mauran25,
  author       = {Mauran, Cecily},
  title        = {Mark Zuckerberg wants {AI} to do half of {Meta}'s coding by 2026},
  organization = {Mashable},
  year         = {2025},
  url          = {https://mashable.com/article/llamacon-mark-zuckerberg-ai-writes-meta-code},
}

@misc{murtz25,
    author       = {Christian Mürtz and Mark Niklas Müller},
    title        = {Agents in the Wild - Dashboard},
    year         = {2025},
    doi          = {10.5281/zenodo.15846865},
    url          = {https://doi.org/10.5281/zenodo.15846865},
    note         = {Interactive web dashboard. Code available at \url{https://github.com/logic-star-ai/insights}},
    howpublished = {\url{https://insights.logicstar.ai}}
}

@online{novet25,
  author       = {Novet, Jordan and Vanian, Jonathan},
  title        = {Satya Nadella says as much as 30\% of {Microsoft} code is written by {AI}},
  organization = {CNBC},
  year         = {2025},
  url          = {https://www.cnbc.com/2025/04/29/satya-nadella-says-as-much-as-30percent-of-microsoft-code-is-written-by-ai.html},
}

@online{palmer25,
  author       = {Palmer, Annie},
  title        = {Shopify CEO: Prove {AI} can't do jobs before asking for more headcount},
  organization = {CNBC},
  year         = {2025},
  url          = {https://www.cnbc.com/2025/04/07/shopify-ceo-prove-ai-cant-do-jobs-before-asking-for-more-headcount.html},
}

@inproceedings{paradis25,
  title={How much does AI impact development speed? An enterprise-based randomized controlled trial},
  author={Paradis, Elise and Grey, Kate and Madison, Quinn and Nam, Daye and Macvean, Andrew and Meimand, Vahid and Zhang, Nan and Ferrari-Church, Ben and Chandra, Satish},
  booktitle={2025 IEEE/ACM 47th International Conference on Software Engineering: Software Engineering in Practice (ICSE-SEIP)},
  year={2025},
}

@article{paul25,
  title={Investigating The Smells of LLM Generated Code},
  author={Paul, Debalina Ghosh and Zhu, Hong and Bayley, Ian},
  journal={arXiv preprint arXiv:2510.03029},
  year={2025}
}

@online{roychoudhary25,
  author       = {Roy Choudhary, Shauvik and Mahajan, Sonal and Bond, Will and Wang, Joseph and Utture, Akshay},
  title        = {{uReview}: Scalable, Trustworthy {GenAI} for Code Review at {Uber}},
  organization = {Uber},
  year         = {2025},
  url          = {https://www.uber.com/us/en/blog/ureview/},
}

@online{shibu25,
  author       = {Shibu, Sherin},
  editor       = {Malamut, Melissa},
  title        = {{Duolingo} Launches 148 AI-Written Courses, Replacing Humans},
  organization = {Entrepreneur},
  year         = {2025},
  url          = {https://www.entrepreneur.com/business-news/duolingo-will-replace-contract-workers-with-ai-ceo-says/490812},
}

@inproceedings{shihab25,
  title={The Effects of GitHub Copilot on Computing Students' Programming Effectiveness, Efficiency, and Processes in Brownfield Coding Tasks},
  author={Shihab, Md Istiak Hossain and Hundhausen, Christopher and Tariq, Ahsun and Haque, Summit and Qiao, Yunhan and Mulanda, Brian Wise},
  booktitle={Proceedings of the 2025 ACM Conference on International Computing Education Research V. 1},
  pages={407--420},
  year={2025}
}

@online{stewart25,
  author       = {Stewart, Ashley},
  title        = {Microsoft Internal Memo: `Using {AI} Is No Longer Optional.'},
  organization = {Business Insider},
  year         = {2025},
  url          = {https://www.businessinsider.com/microsoft-internal-memo-using-ai-no-longer-optional-github-copilot-2025-6},
}

@article{watanabe25,
  title={On the use of agentic coding: An empirical study of pull requests on github},
  author={Watanabe, Miku and Li, Hao and Kashiwa, Yutaro and Reid, Brittany and Iida, Hajimu and Hassan, Ahmed E},
  journal={ACM Transactions on Software Engineering and Methodology},
  year={2025},
  publisher={ACM New York, NY}
}

@article{xu25,
  title={AI-Assisted Programming Decreases the Productivity of Experienced Developers by Increasing the Technical Debt and Maintenance Burden},
  author={Xu, Feiyang and Medappa, Poonacha K and Tunc, Murat M and Vroegindeweij, Martijn and Fransoo, Jan C},
  journal={arXiv preprint arXiv:2510.10165},
  year={2025}
}

@inproceedings{zeng25,
  title={A first look at conventional commits classification},
  author={Zeng, Qunhong and Zhang, Yuxia and Qiu, Zhiqing and Liu, Hui},
  booktitle=ICSE,
  year={2025},
}

@inproceedings{ambati24,
  title={Navigating (in) security of AI-generated code},
  author={Ambati, Sri Haritha and Ridley, Norah and Branca, Enrico and Stakhanova, Natalia},
  booktitle={2024 IEEE international conference on cyber security and resilience (CSR)},
  pages={1--8},
  year={2024},
  organization={IEEE}
}

@article{chretien24,
  title={Impact of AI-tooling on the Engineering Workspace},
  author={Chretien, Lena and Albarran, Nikolas},
  journal={arXiv preprint arXiv:2406.07683},
  year={2024}
}

@inproceedings{jimenez24,
    title={{SWE}-bench: Can Language Models Resolve Real-world Github Issues?},
    author={Carlos E Jimenez and John Yang and Alexander Wettig and Shunyu Yao and Kexin Pei and Ofir Press and Karthik R Narasimhan},
    booktitle={The Twelfth International Conference on Learning Representations},
    year={2024}
}

@inproceedings{li24b,
  title={Assessing the performance of ai-generated code: A case study on github copilot},
  author={Li, Shuang and Cheng, Yuntao and Chen, Jinfu and Xuan, Jifeng and He, Sen and Shang, Weiyi},
  booktitle=ISSRE,
  pages={216--227},
  year={2024},
  organization={IEEE}
}

@article{ng24,
  title={Harnessing the potential of Gen-AI coding assistants in public sector software development},
  author={Ng, Kevin KB and Fauzi, Liyana and Leow, Leon and Ng, Jaren},
  journal={arXiv preprint arXiv:2409.17434},
  year={2024}
}

@article{pandey24,
  title={Transforming software development: Evaluating the efficiency and challenges of github copilot in real-world projects},
  author={Pandey, Ruchika and Singh, Prabhat and Wei, Raymond and Shankar, Shaila},
  journal={arXiv preprint arXiv:2406.17910},
  year={2024}
}

@article{song24,
  title={The impact of generative AI on collaborative open-source software development: Evidence from GitHub Copilot},
  author={Song, Fangchen and Agarwal, Ashish and Wen, Wen},
  journal={arXiv preprint arXiv:2410.02091},
  year={2024}
}

@article{asare23,
  title={Is github’s copilot as bad as humans at introducing vulnerabilities in code?},
  author={Asare, Owura and Nagappan, Meiyappan and Asokan, Nirmal},
  journal=EMSE,
  volume={28},
  number={6},
  pages={129},
  year={2023},
  publisher={Springer}
}

@article{peng23,
  title={The impact of ai on developer productivity: Evidence from github copilot},
  author={Peng, Sida and Kalliamvakou, Eirini and Cihon, Peter and Demirer, Mert},
  journal={arXiv preprint arXiv:2302.06590},
  year={2023}
}

@inproceedings{perry23,
  title={Do users write more insecure code with ai assistants?},
  author={Perry, Neil and Srivastava, Megha and Kumar, Deepak and Boneh, Dan},
  booktitle={Proceedings of the 2023 ACM SIGSAC conference on computer and communications security},
  year={2023}
}

@inproceedings{imai22,
  title={Is github copilot a substitute for human pair-programming? an empirical study},
  author={Imai, Saki},
  booktitle={Proceedings of the ACM/IEEE 44th International Conference on Software Engineering: Companion Proceedings},
  year={2022}
}

@inproceedings{nguyen22,
  title={An empirical evaluation of GitHub copilot's code suggestions},
  author={Nguyen, Nhan and Nadi, Sarah},
  booktitle=MSR,
  pages={1--5},
  year={2022}
}

@inproceedings{pearce22,
  title={Asleep at the Keyboard? Assessing the Security of GitHub Copilot’s Code Contributions},
  author={Pearce, Hammond and Ahmad, Baleegh and Tan, Benjamin and Dolan-Gavitt, Brendan and Karri, Ramesh},
  booktitle={2022 IEEE Symposium on Security and Privacy (SP)},
  pages={754--768},
  year={2022},
  organization={IEEE Computer Society}
}

@inproceedings{siddiq22,
  title={An empirical study of code smells in transformer-based code generation techniques},
  author={Siddiq, Mohammed Latif and Majumder, Shafayat H and Mim, Maisha R and Jajodia, Sourov and Santos, Joanna CS},
  booktitle={2022 IEEE 22nd International Working Conference on Source Code Analysis and Manipulation (SCAM)},
  year={2022},
  organization={IEEE}
}

@inproceedings{vaithilingam22,
  title={Expectation vs. experience: Evaluating the usability of code generation tools powered by large language models},
  author={Vaithilingam, Priyan and Zhang, Tianyi and Glassman, Elena L},
  booktitle={CHI conference on human factors in computing systems extended abstracts},
  pages={1--7},
  year={2022}
}

@inproceedings{yetistiren22,
  title={Assessing the quality of GitHub copilot’s code generation},
  author={Yetistiren, Burak and Ozsoy, Isik and Tuzun, Eray},
  booktitle={Proceedings of the 18th international conference on predictive models and data analytics in software engineering},
  pages={62--71},
  year={2022}
}

@inproceedings{ziegler22,
  title={Productivity assessment of neural code completion},
  author={Ziegler, Albert and Kalliamvakou, Eirini and Li, X Alice and Rice, Andrew and Rifkin, Devon and Simister, Shawn and Sittampalam, Ganesh and Aftandilian, Edward},
  booktitle={Proceedings of the 6th ACM SIGPLAN international symposium on machine programming},
  pages={21--29},
  year={2022}
}

@article{chen21,
  title={Evaluating large language models trained on code},
  author={Mark Chen and Jerry Tworek and Heewoo Jun and Qiming Yuan and Henrique Ponde de Oliveira Pinto and Jared Kaplan and Harri Edwards and Yuri Burda and Nicholas Joseph and Greg Brockman and Alex Ray and Raul Puri and Gretchen Krueger and Michael Petrov and Heidy Khlaaf and Girish Sastry and Pamela Mishkin and Brooke Chan and Scott Gray and Nick Ryder and Mikhail Pavlov and Alethea Power and Lukasz Kaiser and Mohammad Bavarian and Clemens Winter and Philippe Tillet and Felipe Petroski Such and Dave Cummings and Matthias Plappert and Fotios Chantzis and Elizabeth Barnes and Ariel Herbert-Voss and William Hebgen Guss and Alex Nichol and Alex Paino and Nikolas Tezak and Jie Tang and Igor Babuschkin and Suchir Balaji and Shantanu Jain and William Saunders and Christopher Hesse and Andrew N. Carr and Jan Leike and Josh Achiam and Vedant Misra and Evan Morikawa and Alec Radford and Matthew Knight and Miles Brundage and Mira Murati and Katie Mayer and Peter Welinder and Bob McGrew and Dario Amodei and Sam McCandlish and Ilya Sutskever and Wojciech Zaremba},
  journal={arXiv preprint arXiv:2107.03374},
  year={2021}
}

@article{wang21,
  title={Large-scale intent analysis for identifying large-review-effort code changes},
  author={Wang, Song and Bansal, Chetan and Nagappan, Nachiappan},
  journal={Information and Software Technology},
  volume={130},
  pages={106408},
  year={2021},
  publisher={Elsevier}
}

@article{vanderweele19,
  title={Principles of confounder selection},
  author={VanderWeele, Tyler J},
  journal={European journal of epidemiology},
  volume={34},
  number={3},
  pages={211--219},
  year={2019},
  publisher={Springer}
}

@book{varga18,
  title={Unraveling Software Maintenance and Evolution},
  author={Varga, Ervin},
  year={2018},
  publisher={Springer}
}

@article{hartig16,
  title={DHARMa: residual diagnostics for hierarchical (multi-level/mixed) regression models},
  author={Hartig, Florian},
  journal={CRAN: contributed packages},
  year={2016},
  publisher={The R Foundation}
}

@article{mcintosh16,
  title={An empirical study of the impact of modern code review practices on software quality},
  author={McIntosh, Shane and Kamei, Yasutaka and Adams, Bram and Hassan, Ahmed E},
  journal={Empirical Software Engineering},
  volume={21},
  number={5},
  year={2016},
  publisher={Springer}
}

@article{harrison14,
  title={Using observation-level random effects to model overdispersion in count data in ecology and evolution},
  author={Harrison, Xavier A},
  journal={PeerJ},
  volume={2},
  pages={e616},
  year={2014},
  publisher={PeerJ Inc.}
}

@article{zhou11,
  title={Competing risks regression for stratified data},
  author={Zhou, Bingqing and Latouche, Aurelien and Rocha, Vanderson and Fine, Jason},
  journal={Biometrics},
  volume={67},
  number={2},
  pages={661--670},
  year={2011},
  publisher={Oxford University Press}
}

@article{bolker09,
  title={Generalized linear mixed models: a practical guide for ecology and evolution},
  author={Bolker, Benjamin M and Brooks, Mollie E and Clark, Connie J and Geange, Shane W and Poulsen, John R and Stevens, M Henry H and White, Jada-Simone S},
  journal={Trends in ecology \& evolution},
  volume={24},
  number={3},
  pages={127--135},
  year={2009},
  publisher={Elsevier}
}

@inproceedings{hindle08,
  title={What do large commits tell us? a taxonomical study of large commits},
  author={Hindle, Abram and German, Daniel M and Holt, Ric},
  booktitle=MSR,
  year={2008}
}

@book{mcculloch08,
  title={Generalized, linear, and mixed models},
  author={McCulloch, Charles E and Searle, Shayle R and Neuhaus, John M},
  year={2008},
  publisher={John Wiley \& Sons}
}

@article{jenkins05,
  title={Survival analysis},
  author={Jenkins, Stephen P},
  journal={Unpublished manuscript, Institute for Social and Economic Research, University of Essex, Colchester, UK},
  volume={42},
  pages={54--56},
  year={2005},
  publisher={Citeseer}
}

@article{hartung01,
  title={A refined method for the meta-analysis of controlled clinical trials with binary outcome},
  author={Hartung, Joachim and Knapp, Guido},
  journal={Statistics in medicine},
  volume={20},
  number={24},
  pages={3875--3889},
  year={2001},
  publisher={Wiley Online Library}
}

@article{fine99,
  title={A proportional hazards model for the subdistribution of a competing risk},
  author={Fine, Jason P and Gray, Robert J},
  journal={Journal of the American statistical association},
  volume={94},
  number={446},
  pages={496--509},
  year={1999},
  publisher={Taylor \& Francis}
}

@article{tibshirani96,
  title={Regression shrinkage and selection via the lasso},
  author={Tibshirani, Robert},
  journal={Journal of the Royal Statistical Society Series B: Statistical Methodology},
  volume={58},
  number={1},
  pages={267--288},
  year={1996},
  publisher={Oxford University Press}
}

@article{peduzzi95,
  title={Importance of events per independent variable in proportional hazards regression analysis II. Accuracy and precision of regression estimates},
  author={Peduzzi, Peter and Concato, John and Feinstein, Alvan R and Holford, Theodore R},
  journal={Journal of clinical epidemiology},
  volume={48},
  number={12},
  pages={1503--1510},
  year={1995},
  publisher={Elsevier}
}

@article{grambsch94,
  title={Proportional hazards tests and diagnostics based on weighted residuals},
  author={Grambsch, Patricia M and Therneau, Terry M},
  journal={Biometrika},
  year={1994},
  publisher={Oxford University Press}
}

@book{cox84,
  title={Analysis of survival data},
  author={Cox, David Roxbee and Oakes, David},
  year={1984},
  publisher={CRC press}
}

@article{rao81,
  title={The analysis of categorical data from complex sample surveys: chi-squared tests for goodness of fit and independence in two-way tables},
  author={Rao, Jon NK and Scott, Alastair J},
  journal={Journal of the American statistical association},
  volume={76},
  number={374},
  pages={221--230},
  year={1981},
  publisher={Taylor \& Francis}
}

@inproceedings{swanson76,
  title={The dimensions of maintenance},
  author={Swanson, E Burton},
  booktitle=ICSE,
  year={1976}
}

@article{cohen60,
  title={A coefficient of agreement for nominal scales},
  author={Cohen, Jacob},
  journal={Educational and psychological measurement},
  volume={20},
  number={1},
  pages={37--46},
  year={1960},
  publisher={Sage Publications Sage CA: Thousand Oaks, CA}
}

@article{kaplan58,
  title={Nonparametric estimation from incomplete observations},
  author={Kaplan, Edward L and Meier, Paul},
  journal={Journal of the American statistical association},
  year={1958},
  publisher={Taylor \& Francis}
}

\end{document}